\documentclass[prl,onecolumn,superscriptaddress,showpacs]{revtex4}
\usepackage{amsmath,amssymb,amsfonts,bbm,graphicx}
\usepackage{epstopdf}
\newcommand{\ket}[1]{\vert #1 \rangle}

\newcommand{\ketbra}[2]{\vert #1 \rangle \! \langle #2 \vert}
\newcommand{\sandwich}[3]{\left \langle #1 | #2 | #3 \right \rangle}

\newcommand{\average}[1]{\left \langle #1  \right \rangle}
\newcommand{\be}{ \begin{equation} }
\newcommand{\ee}{\end{equation}}
\newcommand{\bae}{\begin{eqnarray}} \newcommand{\eae}{\end{eqnarray}}

\begin{document}

\title{Interplay between coherence and decoherence in LHCII photosynthetic complex}

\author{Paolo Giorda}
\affiliation{ISI Foundation, I-10133 Torino, Italy}
\author{Silvano Garnerone}
\affiliation{Department of Physics and Astronomy and Center for Quantum Information Science \& Technology,University of Southern California, Los Angeles, CA 90089}
\author{Paolo Zanardi}
\affiliation{Department of Physics and Astronomy and Center for Quantum Information Science \& Technology,University of Southern California, Los Angeles, CA 90089}
\author{Seth Lloyd}
\affiliation{Massachusetts Institute of Technology � Research Lab of Electronics and Dept. of Mechanical Engineering 77 Massachusetts Avenue, Cambridge, MA 02139, USA}

\maketitle

\section{Abstract}

This paper investigates the dynamics of excitonic transport in
photocomplex LHCII, the primary component of the photosynthetic
apparatus in green plants.
The dynamics exhibits a strong interplay between coherent processes
mediated by the excitonic Hamiltonian, and incoherent processes due
to interactions with the environment. The spreading of the exciton over
a single monomer is well described by a proper measure of delocalization
that allows one to identify two relevant time scales.  An exciton initially localized
in one chromophore first spreads coherently to neighboring chromophores.
During this initial coherent spreading, quantum effects such as
entanglement play a role. As the effects of a decohering
environment come into play, coherence and decoherence interact
to give rise to efficient and robust excitonic transport, reaching
a maximum efficiency at the levels of decoherence found in
physiological conditions.  We analyze the efficiency for different possible topologies
(monomer, dimer, trimer, tetramer) and show how the trimer has a particular role
both in the antenna and the wire configuration.

\section{Introduction}

Recent experiments probing excitonic transport in green sulphur bacteria
suggest that quantum coherence plays an important role in photosynthesis
\cite{Engel07, Scholes10, Engel10}.  Detailed models of the interplay between
coherent exciton dynamics and decoherence and relaxation induced by
the exciton's environment show that the resulting transport is
robust and efficient \cite{SethFMO1, SethFMO2, Plenio08},
an effect known as environmentally
assisted quantum transport (ENAQT).
This paper extends these analyses to the
light-harvesting complex of green plants \cite{IshiFleming10}, specifically, transport
through the light-harvesting complex LHCII .  Excitonic transport
through sets of coupled LHCII complexes \cite{Grondelle06,Exciton09,Fleming09,RengerLHCIIsinks10} differs in significant
ways from the transport through the Fenna-Matthews-Olson complex
(FMO) of green bacteria .
Notably,  the LHCII can act both as antennae and 'wires'
capable of transferring the excitons captured by other
complexes through the structure.  In order to move through a
sequence of LHCII complexes, the exciton must move both up
and down in energy, a process mediated by interactions with
the environment.
 LHCII complexes can be found in different forms allowing for
various regulation activities. In particular, some LHCII complexes can migrate, under specific
light conditions,  from the Photosystem II (PSII) to Photosystem I (PSI) in order to optimize the
photosynthetic process (state transition), or disassamble into monomeric subunits in order
to favor the regulation of light harvesting in excess light \cite{LHCIItrimervsmonomer}.

This paper applies a purely-decohering
Haken-Strobl model to analyze interactions between coherent
and decoherent dynamics in excitonic transport in LHCII
\cite{HakenStrobl, HakenReineker}.
The advantage of using Haken-Strobl model is that it is
the simplest model that allows
this transport to be investigated in the strongly-coupled,
non-perturbative regime.  The disadvantage of
a purely decohering model is that it does not
include relaxation, and so will over-estimate
transport rates as the exciton moves up in energy, and
under-estimate them as it moves down.  Nonetheless, our
analysis shows that pure decoherence is a surprisingly
effective transport mechanism even when the exciton
is moving down through an energy funnel.

To analyze the interplay of coherence and decoherence, we use
familiar tools from quantum information theory.  Quantum mutual
information between sets of sites is used to track the pathways by which
correlations spread through the LHCII monomeric complex due to exciton motion.
Negativity and concurrence are used to demonstrate and quantify
entanglement.  The comparison among the different measures allows us
to identify
two timescales.  Over the first half picosecond, an initially localized
exciton spreads coherently to neighboring chromophore.  This coherent
spreading is accompanied by rapid oscillations in quantum mutual
information and negativity, indicating the presence of significant
coherence and entanglement.   Next, over the course of several
picoseconds, the coherent oscillations disappear, the negativity
decreases, and quantum mutual information between sites grows as
decoherence kicks in and the exciton diffuses partially incoherently
to more distant sites.  The interplay between coherence and decoherence
gives rise to highly efficient transport through the complex.
We analyze the efficiency of transport for various geometries
of LCHII (monomer, dimer, trimer, and tetramer) and  for the two possible configurations: antenna and wire.
We show that efficiency increases as the number of subunits increases,
saturating at the trimer.  That is, the dimer exhibits more efficient
transport than the monomer; the trimer is more efficient than
the dimer; and the trimer and tetramer give the same efficiency.

\section{Background}

Photoabsorption, the initial step of photosynthesis, takes place in photosynthetic complexes
formed by groups of pigments (chlorophylls, Chl) and proteins placed within the thylakoid membrane
\cite{Photosynthesisbook}. The light-harvesting pigments are arranged in protein matrices in such a way
that the photo-excitation is funneled to the reaction center, where the energy carried by the excitation is
used for fueling chemical reactions.

One of the main elements in the photosynthetic complexes of higher plants is
the light-harvesting complex LHCII. In the following we will refer to the LHCII that can be found in the
{\it Arabidopsis thaliana}, which is a flowering plant that has been extensively studied as a model in
plant biology and genomics.
In this plants, LHCII can be found as peripheral antenna in the supercomplex
PSII \cite{LHCIIwebsite}. Its role in the supercomplex is to
collect photons from the incoming light or excitons from the neighbouring complexes (other LHCIIs or other
minor antenna complexes) and transfer them to other LHCII complexes or to the reaction center.

\begin{figure}
\fbox{
\includegraphics[width=9cm,clip]{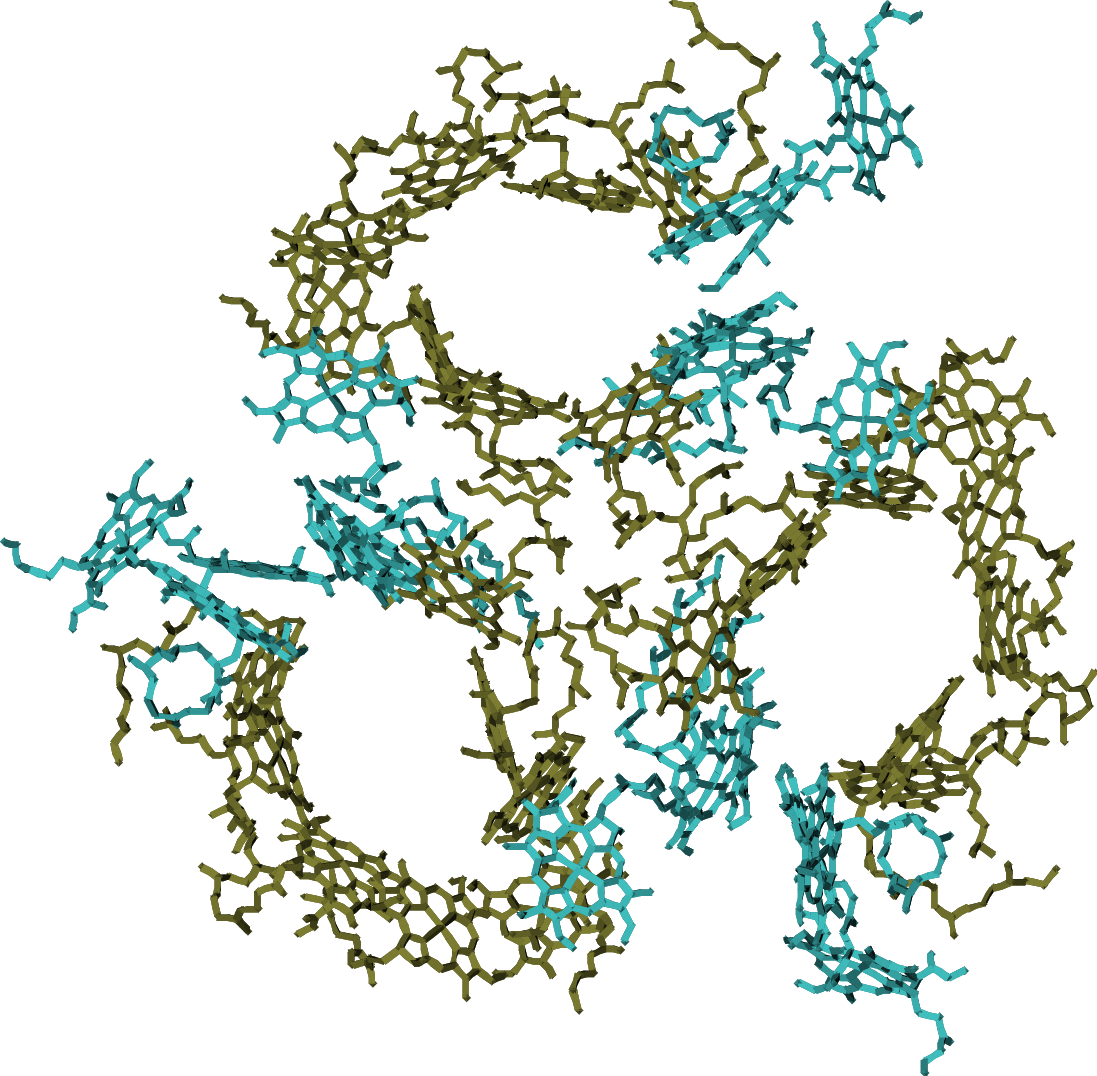}
}
\caption{The chains of Chls in the trimeric LHCII. Blue: CHls-b; green: CHls-a. }
\label{Fig.: trimer}
\end{figure}

\begin{figure}
\fbox{
\includegraphics[width=9cm,clip]{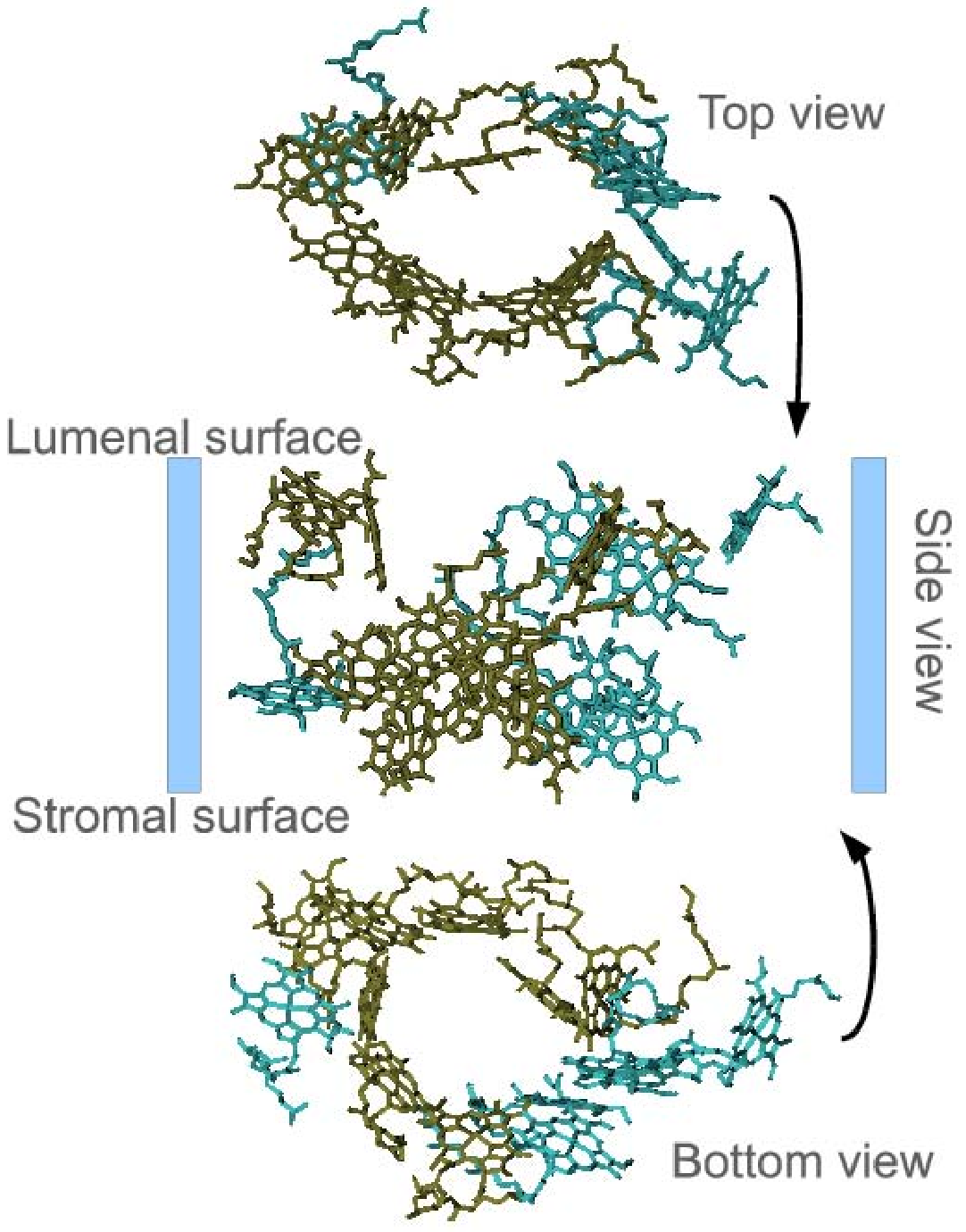}
}
\caption{LHCII monomer in the membrane, and its top and bottom view. Blue: CHls-b; green: CHls-a}
\label{Fig.: strlumChls1}
\end{figure}

The LHCII crystal structure has been determined up to $2.5\, {\AA}$ in resolution
\cite{LHCIIcrystal1, NatureSpinach}:
it is composed of three similar monomeric subunits (Lhcb1-3) each containing
14 Chls molecules embedded in a protein matrix (see Fig.\ref{Fig.: trimer}). Two different types of Chls are present: $8$ Chl-a type,
and $6$ Chl-b type. The main difference between the two is the presence in Chl-b of a carbonyl group that
allows for higher excitation energies. The Chls are disposed in two layers which are called {\it stromal}
and {\it lumenal}, the first being oriented towards the outer part of the thylakoid membrane, the second
towards its inner part (see fig. \ref{Fig.: strlumChls1}).

\begin{figure}
\fbox{
\includegraphics[width=5cm]{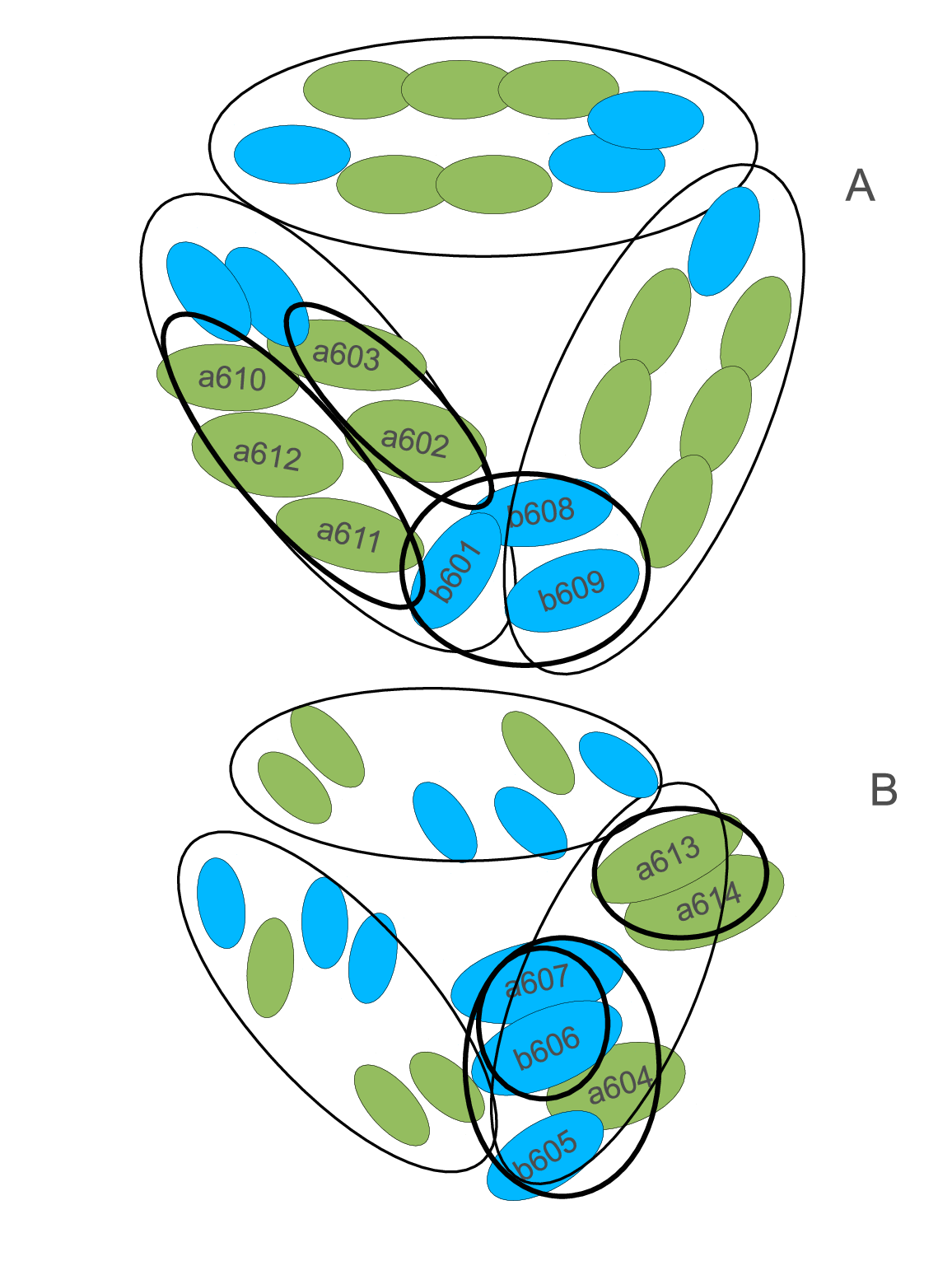}
}
\caption{Stromal (A) and lumenal (B) layers of the LHCII. The groups of strongly coupled Chls are enclosed in
thick circles.}
\label{Fig.: strlumChls2}
\end{figure}

Within each layer, the Chls can be grouped on the basis of their relative distance and orientations which
determine the strength of the electronic interaction. The main groups of Chls are depicted in fig.
\ref{Fig.: strlumChls2}; different groups have different linear absorption spectra and two main bands can be
identified: the Chl-a band centered at $14925\, cm^{-1}$, and the Chl-b band centered at
$15385\, cm^{-1}$.

The exciton energies and the relative pigment participations
in LHCII have been determined in \cite{Exciton09}.
The energy relaxation pathways of the system were described by van Grondelle (see
\cite{Grondelle06}) and refined in  \cite{Fleming09} by means of the
experimental data obtained by applying 2D spectroscopic techniques to samples of LHCII at $77 K$; in the
same paper, the Hamiltonian for the monomeric unit we will use in the following was also derived  and
optimized in order to account for the experimental results.
The Hamiltonian is written in the site basis: the diagonal terms are the site energies, determined by
fitting the linear absorptions spectrum  and the off-diagonal terms account for the Coulomb interactions
between pairs of Chls (see \cite{Fleming09} and references therein).
The experimental evidences allows to identify two main downhill relaxation pathways
\cite{Grondelle06,Fleming09}. One takes entirely place at the stromal level
(stromal-stromal) while the second starts in the lumenal level (lumenal-stromal) (see fig. \ref{Fig.:
pathways}).  

\begin{figure}
\fbox{\includegraphics[width=9cm,clip]{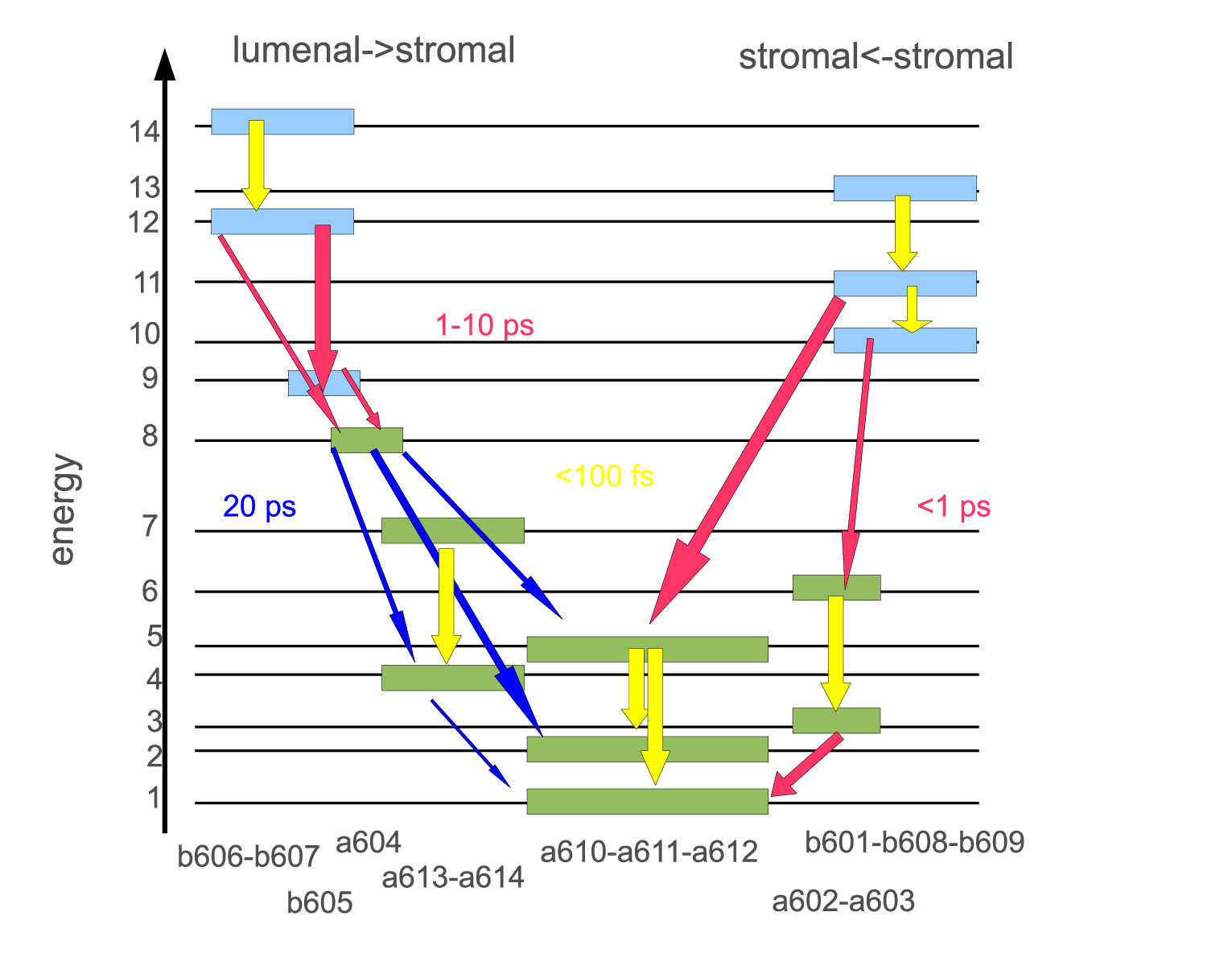}}
\caption{Energy relaxation pathways as experimentally determined in \cite{Fleming09}.}
\label{Fig.: pathways}
\end{figure}

One of the fundamental mechanisms that is at the basis of the energy transfer through the LHCII is given by
the interplay between strong electronic couplings between nearby Chls. The coupling allows both for the
energy splitting between excitonic levels and for the exciton delocalization. If we focus for example on
the group of Chls $a610,a611,a612$ (fig. \ref{Fig.: strlumChls2}) we see that the contiguity and the
relative orientation of these molecules result in a strong interaction which in turn it allows for the
presence of three exciton levels (levels $5,2,1$ in fig. \ref{Fig.: pathways}) which are separated in
energy and at the same time are spatially overlapping.
In presence of exciton-exciton coupling, for example mediated by the
environment, the spatial overlap
allows for fast ($< 100 fs$) relaxation processes within the group-excitonic band (levels $5,2,1$ in fig.
\ref{Fig.: pathways}).

The same mechanism can be seen in another group of Chls $b601,b608,b609$ which allows
for a fast ($< 100 fs$) relaxation process within the stromal b-band (levels $13,11,10$ in fig. \ref{Fig.:
pathways}).

In analogy with what happens for the previous groups of Chls, the intra-group energy pathways can be
described by the same mechanism. In particular, in the lumenal-stromal pathways, one can identify a
bottleneck of the pathway that is given by the Chls $b605$ and $a604$; here the interaction between the
two molecules gives rise to a pair of well localized excitonic states (levels $8$ and $9$ in fig. \ref{Fig.:
pathways}) which have poor spatial overlap with the lower energy states concentrated on the Chl-a groups in
the lumenal and the stromal side. The localization of the wave function gives rise to the experimentally
determined slowdown in the relaxation process relative to the
lumenal-stromal pathway.
%This bottleneck is conjectured to play a role in photoprotection:
%triplet exciton states whose energy is sufficiently high to create potentially
%harmful singlet oxygen can be trapped here and transferred to a carotene
%molecule, where they can be quenched without damage.
The dynamics of the bottleneck will be seen in action in the simulations
of exciton transport through the monomer in the next section.

The lowest energy states of the monomer are two of the excitonic states localized in the $a610,a611,a612$
group of Chls. These sites are therefore indicated as the output sites of the monomer unit (donors), which
can be coupled to other LHCII monomeric units, or other complexes in the PSII. The coupling can be between
exitonic states corresponding to similar energies but located on different neighbouring complexes. In
\cite{Fleming09} these two output states have been studied in terms of their directionality,
which is determined by the interplay between the site basis contribution to the specific excitonic state
(delocalization over the group of Chls $a610,a611,a612$) and the relative orientation of the donor
transition dipoles. The result of the calculations showed that the two excitons can act as donor states in
two different directions, and this mechanism has been suggested as a way that LHCII complexes have to
optimize the exciton transfer to other complexes even in presence of their misalignment.
In our analysis we want to compare two different ways in which the monomer can be coupled with the outer complexes: $i)$ a
sites based coupling, where each output site (Chls $a10,a611,a612$) is independently coupled with an
external complex that we model as a sink; $ii)$ an exciton based coupling, where the two lowest excitonic
states are independently coupled with an outer sink.

For the inter-monomeric couplings, the experimental evidences in \cite{NatureSpinach}
(\cite{Fleming09}), show that the main coupling should be localized in the stromal side
between Chls of b type ($b601,b608,b609$); the strength of the coupling should be of the order of $42
cm^{-1}$, ($35 cm^{-1}$). In \cite{Fleming09} this inter-monomeric coupling has been neglected
in the determination of the single monomer Hamiltonian. While this has the effect of shifting the excitonic
b-band, it should not significantly affect the other transitions which are localized within each monomer. In
order to have a description of the whole trimeric LHCII, we reintroduce the coupling between Chl-b
pertaining to different adjacent monomers; in particular we choose two different configurations:
$b601\leftrightarrow  b609$ and $b601\leftrightarrow b609, b608$.\\
We finally note that another Hamiltonian for the LHCII complex has been derived in \cite{RengerLHCIIsinks10}. However the
results that can be obtained in the following analysis by using the data reported in \cite{RengerLHCIIsinks10} instead of the ones reported in 
\cite{Fleming09}  do not change in a significant way. 

\section{Model and tools}

In the following we will focus on transport properties of the monomeric, dimeric,
trimeric and tetrameric LHCII
systems. We study the dynamics of the open quantum system in the presence of dephasing processes (Haken-
Strobl formalism) in the presence of recombination and trapping mechanisms.
In this model we resort to the description used in \cite{SethFMO1, SethFMO2},
where recombination and trapping
processes are modeled by adding non-Hermitian terms to the Hamiltonian.
The effects of static disorder will be taken into account below.
The equation of motion for the
density matrix of the monomer subunit can be written as
\bae
\frac{d\rho}{dt}&=&\frac{-i}{\hbar}\left[H_{monomer},\rho(t)\right]+L_{deph}\rho(t)-\left\{
H_{recomb}+H_{trapping},\rho\right\} \equiv \nonumber \\
&\equiv& L [\rho(t)],
\eae
The free Hamiltonian of the monomer is a tight binding Hamiltonian and it is expressed in terms of the site
energies and couplings given in \cite{Fleming09}
\be H_{monomer}=\sum_{m}\epsilon_{m}|m\rangle\langle m|+\sum_{m<n}V_{mn}(|m\rangle\langle n|+h.c)
\ee
The term
\be
L_{deph}\rho(t)=\gamma_\phi\sum_{m}A_{m}\rho(t)A_{m}^{\dagger}-\frac{1}{2}\left\{ A_{m}A_{m}^{\dagger},
\rho(t)\right\} ,\ee
accounts for the presence of pure dephasing, $A_{m}=|m\rangle\langle m|$.
The term
\[
H_{recomb}\equiv-i \Gamma\sum_{m}|m\rangle\langle m|,\]
accounts for the recombination processes.
For the trapping, we suppose that once the exciton has reached the output sites
$a610,a611,a612$, it leaves the LHCII complex with a rate $k_{trap}$.
The trapping process can be expressed either with respect to the site basis i.e., the sites
$a610,a611,a612$ are supposed to be singularly linked with other surrounding complexes
\be H_{trapping}\equiv-i k_{trap}\sum_{m=610,611,612}|m\rangle\langle m|,
\label{Eq. : HtrapSites}\ee
or with respect to the two lowest exciton eigenstates that are localized in the
$a610,a611,a612$ sites:
\be H_{trapping}\equiv-i k_{trap}(\ketbra{E_1}{E_1}+\ketbra{E_2}{E_2}).
\label{Eq. : HtrapExcitons}
\ee
The main difference between the two pictures should be that with the site-based trapping mechanism the
sink acts on the output sites,
including the case
where only the highest exciton localized on the output states (number $5$
in figure \ref{Fig.: pathways}) is populated.
By contrast, with the excitonic-based mechanism only the two lowest energy
excitonic states are involved in the trapping dynamics.

The functionals we use to evaluate the efficiency of the transport are the efficiency $\eta$, defined as
\be
\eta=2k_{trap}\sum_{m}\int_{0}^{\infty}dt\langle m|\rho(t)|m\rangle
\label{Eq. : defeta}
\ee
and the average transfer time $\tau$, defined as
\be
\tau=\frac{2}{\eta}k_{trap}\sum_{m}  \int_{0}^{\infty}dt\, t \, \sandwich{m}{\rho(t)}{m}
\label{Eq. : deftau}
\ee

The system dynamics and the efficiency $\eta$ can be evaluated numerically
by vectorizing the density matrix
and constructing the proper linear super operator associated to $L$\[
\rho(t)\rightarrow|\rho(t)\rangle\rangle\]
\[
L\rightarrow\mathcal{L}\]
\[
|\frac{d\rho(t)}{dt}\rangle\rangle=\mathcal{L}|\rho(t)\rangle\rangle\implies|
\rho(t)\rangle\rangle=e^{\mathcal{L}t}|\rho(0)\rangle\rangle.\]
 $\mathcal{L}$ has been constructed using the identity $|ABC\rangle\rangle=A\otimes C^{t}|B\rangle\rangle,$
 where A,B and C are matrices of size n and $|\cdot\rangle\rangle$ is a vector of size $n^{2}.$ We can now
 compute $\eta$ in terms of $\mathcal{L}$ as \[
\eta=-2k\sum_{m}\langle\langle m|\mathcal{L}^{-1}|\rho(0)\rangle\rangle.\]

\begin{figure}
\fbox{\includegraphics[width=9cm,viewport= 16 8 540 480,clip]{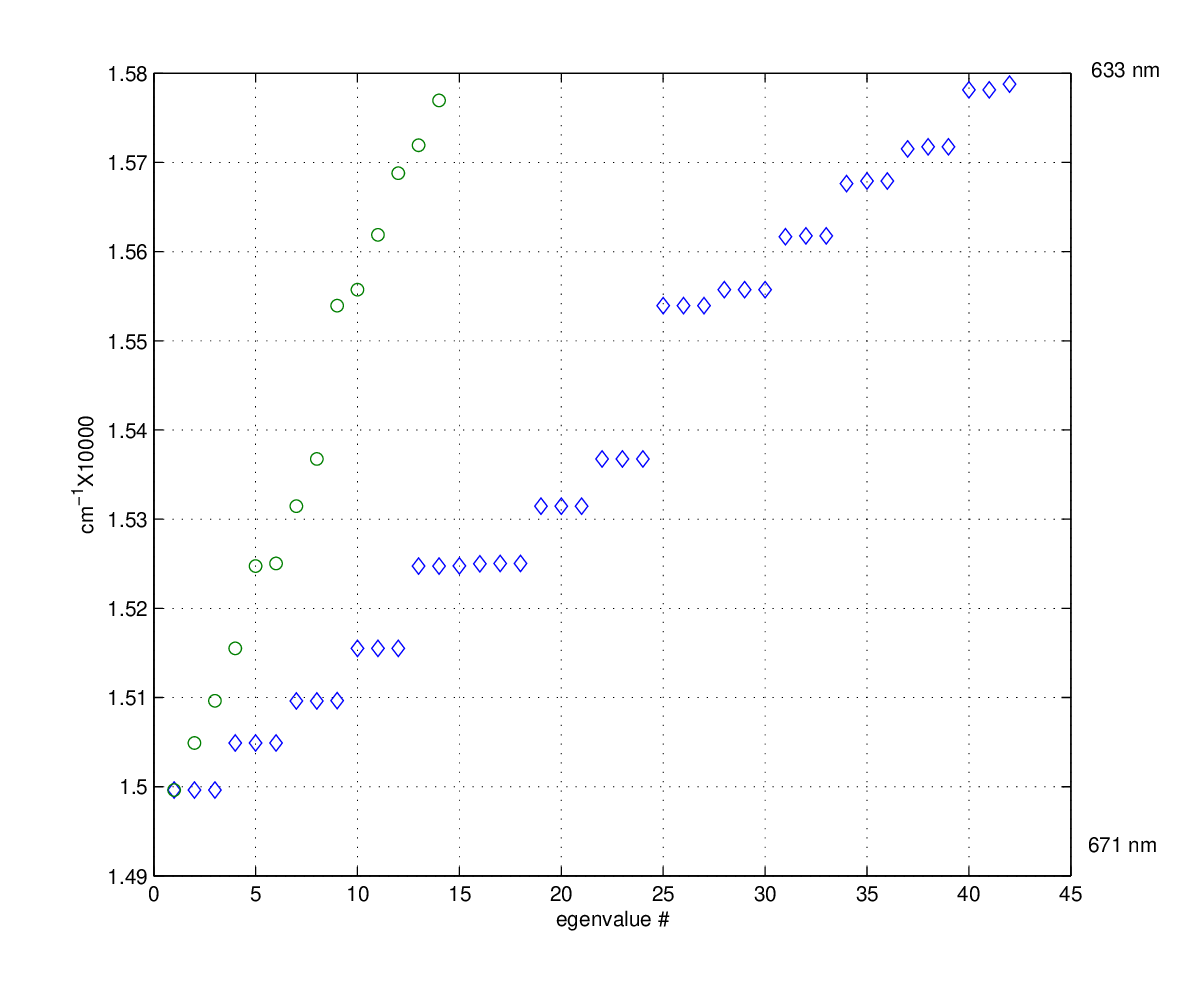}}
\caption{Energy levels for the Hamiltonian of the monomer and the trimer.}
\label{Fig. : HamSpectrum}
\end{figure}

Before passing to analyze the results of our simulations a comment on the monomer Hamiltonian is in order.
The Hamiltonian given in \cite{Fleming09} has been obtained after many optimization processes
with the goal of faithfully reproducing the 2-D spectroscopy experimental results. This implies that the
actual eigenstates that can be derived by diagonalizing the Hamiltonian are delocalized over groups of Chls
which are sometimes different by the ones showed in fig. \ref{Fig.: pathways}, where the relative pigment
participations are derived via other optimization processes. The differences manly involve the highest
energy eigenstates of the b lumenal and stromal band. In our analysis,
we stick to the experimentally optimized
Hamiltonian and use its actual eigenstates; the spectrum of the monomer and the trimer is given
in figure \ref{Fig. : HamSpectrum}.
The maximal, minimal and average difference in energy for the
monomer are $\Delta_{min} \approx 4
cm^{-1},\Delta_{max}\approx 172 cm^{-1} ,\Delta_{Av}= 59 cm^{-1}$ ($k_B T \approx 200 cm^{-1}$ at room
temperature). The differences  in energies for the dimer and the trimer are very similar.

In order to study the dynamics of the monomer we make use of well established quantum information
measures of correlations. In particular, we measure the {\it total} amount of correlations between two subsystems
of chromophores $A$ and $B$ with the quantum mutual information \cite{NielsenChuang}

\be
\mathcal{I}_{AB}= \mathcal(S)_A + \mathcal(S)_B -\mathcal(S)_{AB}
\ee
where $\mathcal(S)_X = \sum_i \lambda_i \log_2{\lambda_i}$ is the von Neumann entropy of the
reduced density matrix $\rho_X$ of subsystem $X$ evaluated in terms of its eigenvalues $\{\lambda_i\}$.
As for the {\it quantum} correlations between subsystems composed by arbitrary number of chromophores
we use the negativity $\mathcal{N}_{AB} $ \cite{vidalvernerneg} that in the
single exciton manifold can be written as \cite{Caruso10}
\be
\mathcal{N}_{AB}= \sqrt{a_{00}^2 + 4 \sum_{n=1}^k \sum_{m=k+1}^N |\sandwich{n}{\rho_{AB}}{m}|^2}-a_{00}
\ee
where $a_{00}$ is the element corresponding to the zero exciton subspace and $A=1,\cdots ,k$, $B=k+1,\cdots,N$
are two generic subsystems of chromophores.
The quantum correlations between pairs of sites $m,n$ are also measured by means of the concurrence \cite{WoottersConcurrence}
which in the single site exciton has the simplified form \cite{Sarovar10}
\be
\mathcal{C}_{m,n}=2|\rho_{m,m}|.
\ee

In order to study the relationship between the dynamics of the above described correlations measures and the
delocalization of the exciton over the monomeric structure we define a measure of delocalization $\mathcal{D}(t)$ that involves the single site populations.
In keeping with this paper's methods of using information-based measures to
characterize the excitonic transport, we use Shannon entropy as a
measure of delocalization.  If $\average{n_i}$ is the
population of site $i$ at time t, then by using the normalized populations $\lambda_i=\average{n_i}/\sum_i
\average{n_i}$ we can define
\be
\mathcal{D}(t)=-\sum_i \lambda_i \log{\lambda_i}
\label{Eq. : Delocal}
\ee
as the Shannon entropy of the normalized populations.
The higher $\mathcal{D}(t)$, the flatter the probability distribution $\{\lambda_i\}$, the higher
the delocalization of the exciton over the complex.

\section{Monomer dynamics}
In the following we describe the dynamics of the monomer. The values of the recombination and
trapping coefficients used for the simulations are the ones used for the FMO complex in
\cite{SethFMO1,IshiFleming10}.
The recombination coefficient $\Gamma=10^{-3}ps^{-1}$ takes into account the
estimated lifetime of the exciton, $1 ns$. The trapping coefficient is $k_{trap}= 1 ps^{-1}$ and is
assumed to be equal for each output exciton state.
Our results do not depend sensitively on the exact value of the exciton
lifetime and the trapping rate: what is important for the analysis
is that the exciton
has a relatively long lifetime compared with coupling rates, and that
the trapping rate is comparable to those rates.

We first focus on the time simulation of the evolution of the monomer in order to identify the possible
energy transfer pathways. We therefore fix the value of the dephasing rate $\gamma_{\phi}\approx 3 ps^{-1}$ that corresponds to $77 K$
(temperature at which the experiments were done in \cite{Fleming09}). The dephasing rate can be
written in terms of the bath correlation function as \cite{SethFMO2}:
\be
\gamma_\phi(T)= 2 \pi\frac{k_B T}{\hbar}\partial_\omega J(\omega)|_{\omega=0}= 2 \pi\frac{k_B T}
{\hbar}\frac{E_r}{\hbar \omega_c}
\label{Eq. : gamma_phi}
\ee
where we have supposed to have an Ohmic correlation function $J(\omega)=\pi\frac{k_B T E_r}
{\hbar}\frac{\omega}{\omega_c}\exp{\omega/\omega_c}$ (super- and sub-Ohmic
correlation functions give a similar dependence on $T$, $E_r$, and $\omega_c$).
The recombination energy $E_r=35 cm^{-1}$ and the cut-off
frequency $\omega_c=150 cm^{-1}$ are chosen to be the
ones used for the FMO simulations.  Again, the qualitative behavior
of the excitonic transport does not depend sensitively on
the precise values of $E_r$ and $\omega_c$.

\subsection{Energy and correlation pathways}
We want to describe the time evolution of the state of the
monomer coupled with the environment. We first try
to identify and characterize the existence of two possible  pathways, stromal-stromal and lumenal-stromal,
by which the exciton, starting from a high energy state belonging to the b band, can reach the output
sites. We therefore focus our attention on the behaviour of the system {\it without any trapping} and for a
value of the dephasing rate that corresponds to the temperature of $77 K$ used in
\cite{Fleming09}.  As noted above, since the Haken-Strobl
model is purely dephasing and does not include an explicit relaxation term,
we expect our analysis to underestimate the rate of transfer from high
energy states to low energy states.  Nonetheless, as will now be seen,
pure decoherence gives efficient excitonic transport down the energy
ladder.

The distinct pathways can be studied by plotting the following populations: $i)$ the populations of the
excitonic states which are mostly localized on Chl-b  molecules that belong to the stromal (lumenal) side
$P_{bStrom} (P_{bLum})$ $ii)$ the populations of the excitons states which are mostly localized on Chl-a
molecules that belong to the stromal (lumenal) side $P_{aStrom} (P_{aLum})$.

\begin{figure}
\fbox{\includegraphics[width=9cm,viewport= 1 220 740 800,clip]{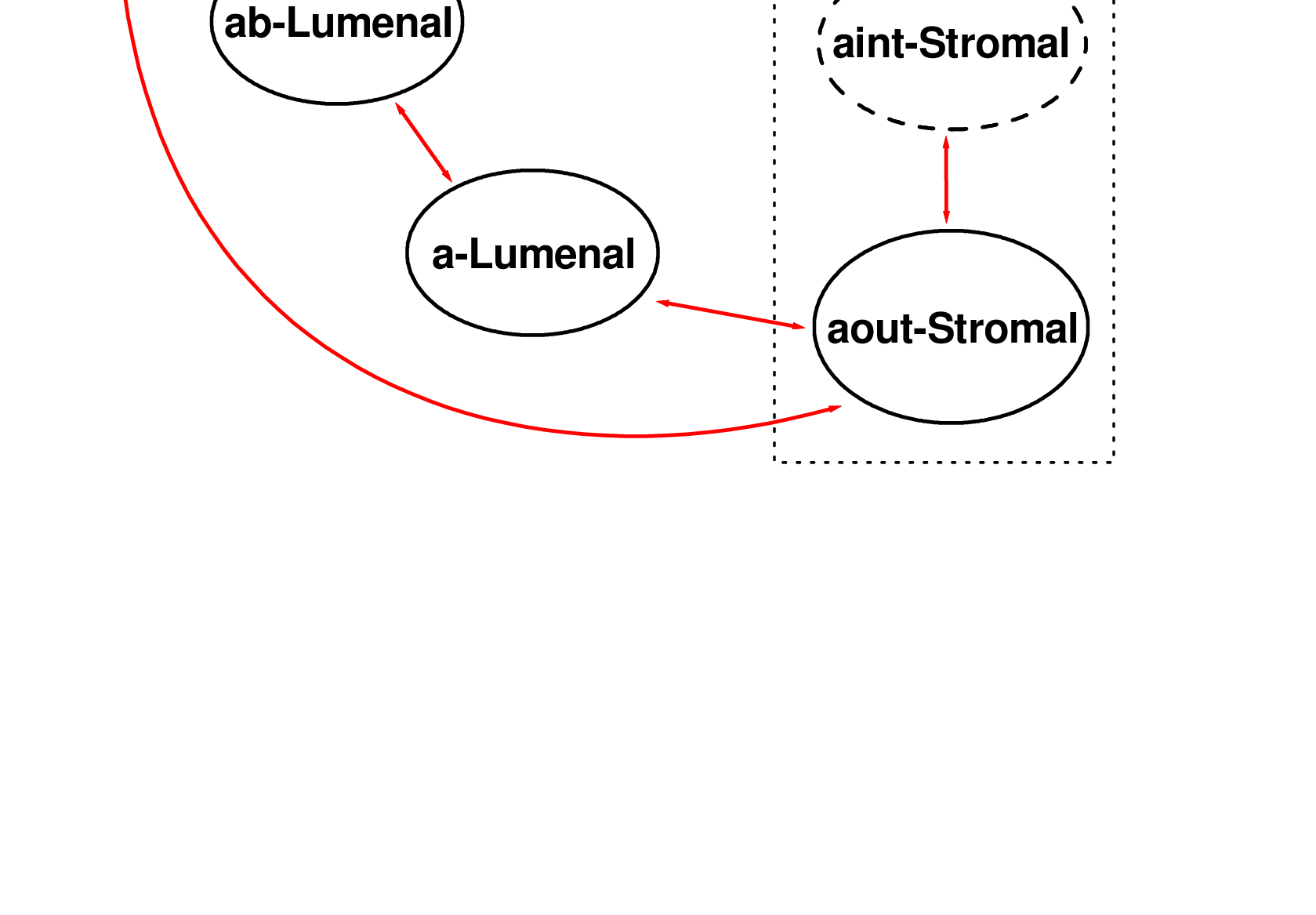}}
\caption{Schematic picture of the bipartitions used for the analysis of correlations dynamics}
\label{Fig.: bipartitions}
\end{figure}

The different excitonic behaviors along the two distinct pathways can also be highlighted by using measures of quantum correlations. In particular,
we study the quantum mutual information between the relevant subsystems that are naturally suggested by the energy landscape in fig. \ref{Fig.: pathways};
in this way we can identify the {\it correlation pathways} and their dynamics.
On the lumenal side we
select the subsystems $bL=\{606,607\},abL=\{605,604\},aL=\{613,614\}$, while on the stromal side we select
the subsystems $bS=\{601,608,609\},aintS=\{602,603\},aoutS=\{610,611,612\}$ and $aS=aintS \cup aoutS$; the
bipartitions are schematically depicted in fig. \ref{Fig.: bipartitions}.
The growth of quantum correlations between subsystems is a signature
of the spreading of the initially localized exciton between subsystems.
The form that these correlations take over time reveals the mechanism
of this spreading -- an almost purely coherent initial propagation
followed by semi-coherent diffusion.

\begin{figure}
\fbox{\includegraphics[width=9cm,viewport= 32 8 500 300,clip]{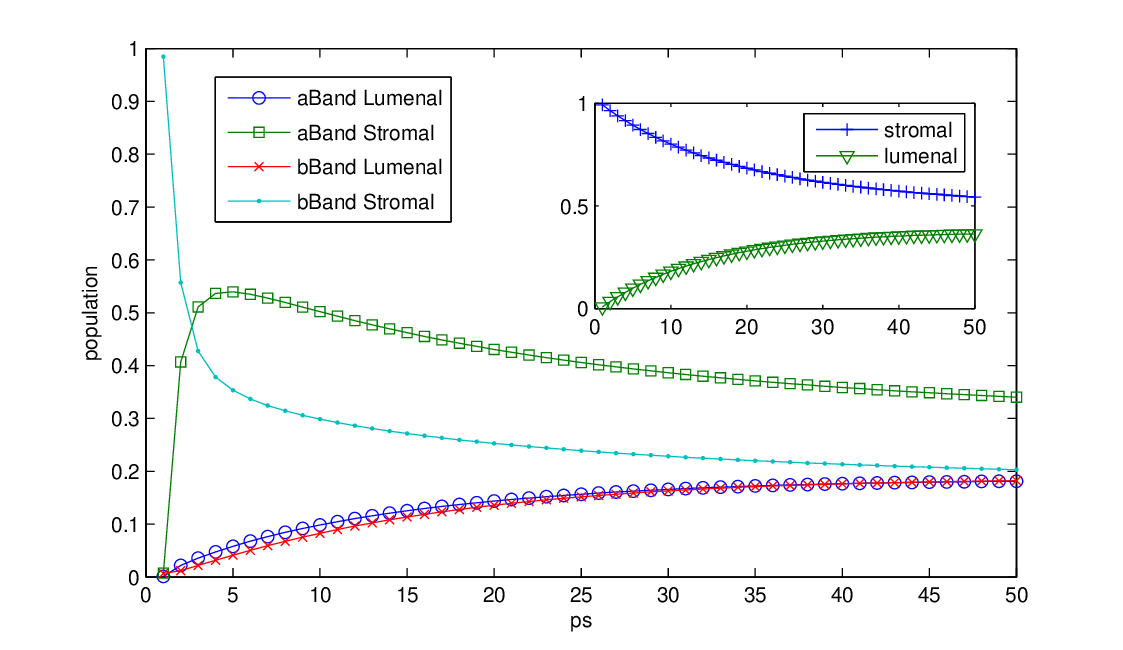}}
\caption{Stromal-Stromal pathway: populations for initial state $\ket{E_{14}}$}
\label{Fig.: FigE14PathSS}
\end{figure}

For the stromal-stromal pathway we choose as initial state of our simulations the highest energy eigenstate
$\ket{E_{14}}$ of the Hamiltonian, which is mostly localized on the $b601$ Chl.
In Fig.\ref{Fig.: FigE14PathSS} we see how the exciton mostly flows from the stromal b-band to the stromal
a-band on a very short time scale ($\approx 5 ps$). On a slower time scale the population partially
delocalizes over the lumenal band. The flow of population between the two b-bands was highlighted in the
energy pathway given in \cite{Grondelle06} (but not highlighted in
\cite{Fleming09}), where the possibility of a flow from the {\it b-lumenal to the b-stromal}
band is estimated to be of the order of $2 ps$. Here we observe the inverse passage {\it b-stromal to b-
lumenal} and this is due also to the partial delocalization of the $\ket{E_{14}}$ on the b-lumenal sites.
The global population decreases because of excitonic decay
with a time scale of the order of $1 ns$.

%\begin{figure}
%\begin{center}
%\includegraphics[width=0.45\textwidth,viewport= 32 8 500 300,clip]{FigE14PathSS}
%\includegraphics[width=0.45\textwidth,viewport= 1 280 640 800,clip]{MIE14Tr0Stromal}
%\caption{Left: Stromal-Stromal pathway: populations for initial state $\ket{E_{14}}$. Right: 3-rd generation RHF with $f=3$} \label{Fig: regulargraphs}
%\end{center}
%\end{figure}

\begin{figure}
\fbox{\includegraphics[width=8cm,viewport= 1 280 640 800,clip]{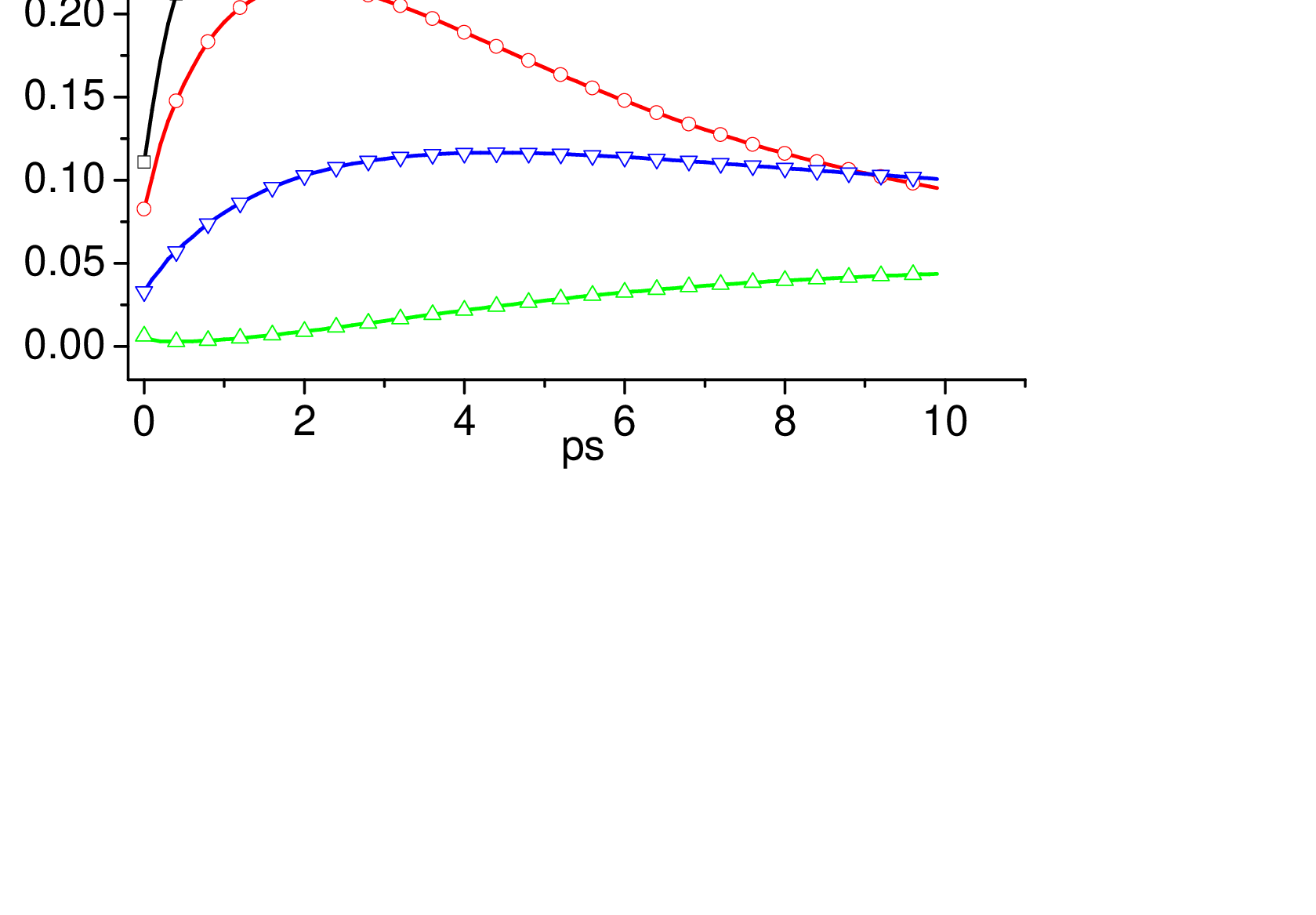}
\includegraphics[width=8cm,viewport= 1 280 640 800,clip]{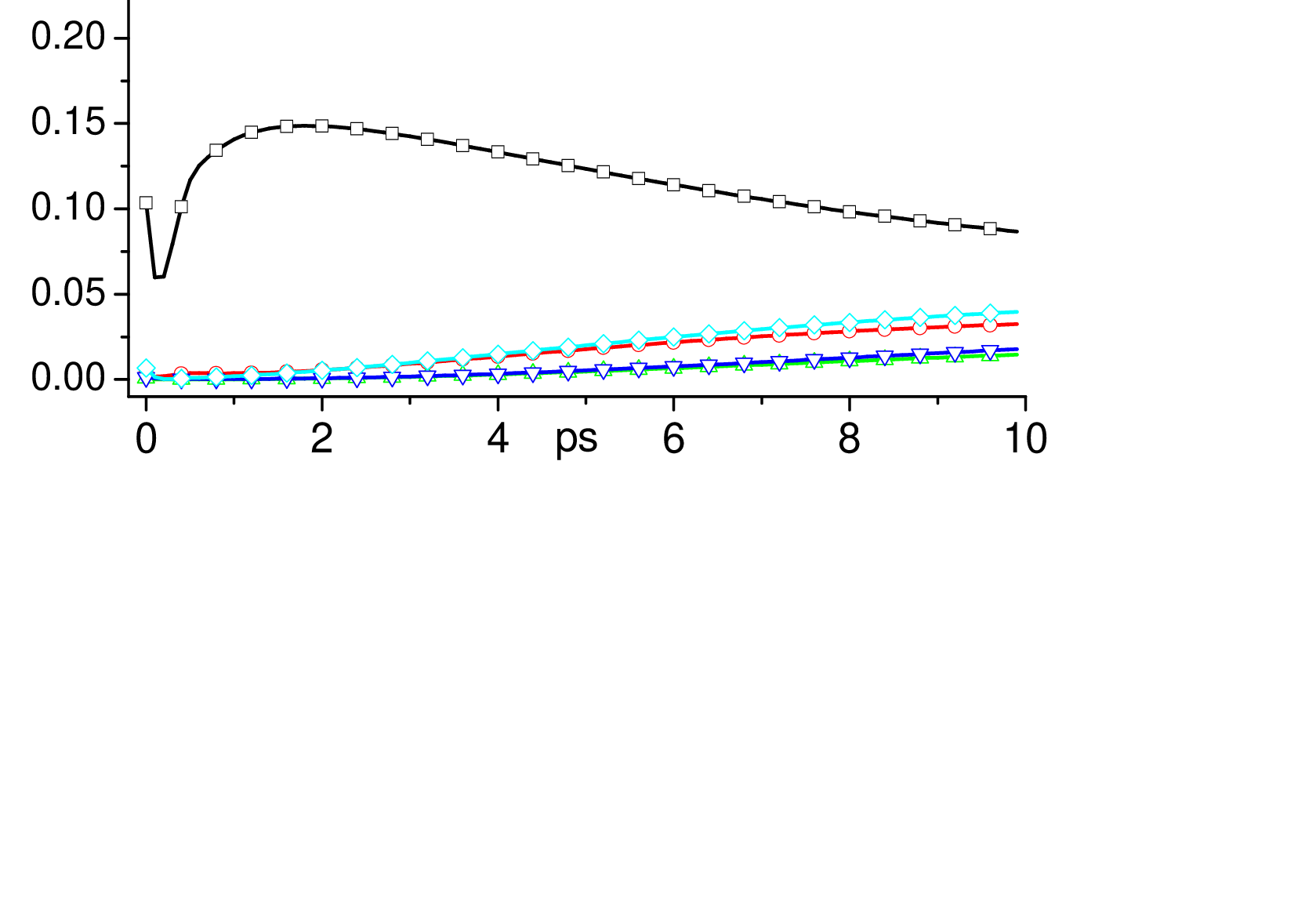}}
\caption{Stromal-Stromal pathway: mutual information for initial state $\ket{E_{14}}$; left: stromal side;
right: lumenal side}
\label{Fig.: MIE14Tr0Deph3}
\end{figure}

The dynamics can be further analyzed by considering the correlation pathways.
Figure \ref{Fig.: MIE14Tr0Deph3} refers to the stromal-stromal pathway starting with $\ket{E_{14}}$. The
plots shows that the dynamics mostly takes place on the stromal side. In particular (left plot) the $bS$
subsystem initially gets correlated with a-band stromal sites as a whole ($MIaSbS$); in the first few
picoseconds most of the correlations are established between the subsystems $bS$-$aintS$, and the
subsystems $aintS$-$aoutS$. From the first picosecond on the $bS$ sites get directly correlated with the
output sites $aoutS$.
The right plot in fig. \ref{Fig.: MIE14Tr0Deph3}  shows that while there are correlations between the
stromal and the lumenal b-bands ($MIbLbS$), the correlations among the subsystems on the lumenal side and
the lumenal-stromal correlations at the level of the a-bands are negligible ($\approx$ one order of
magnitude smaller). This picture is consistent with a dynamics mostly localized on the stromal side of the
monomer.

\begin{figure}
\fbox{\includegraphics[width=9cm,viewport= 32 8 500 300,clip]{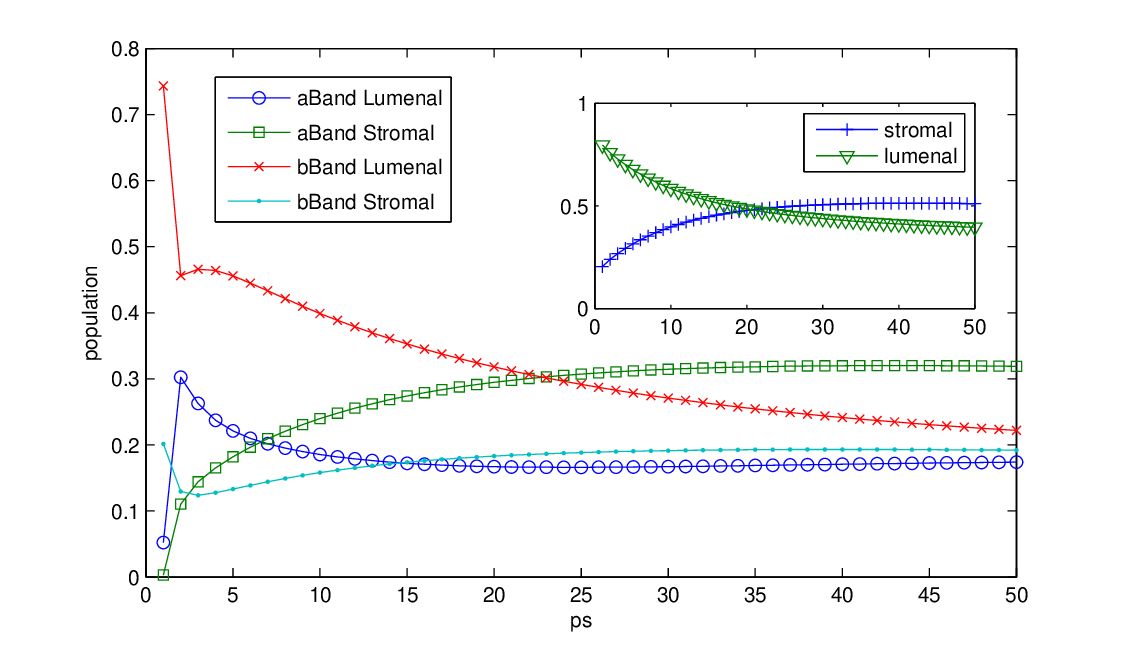}}
\caption{Lumenal-Stromal pathway: populations for initial state $\ket{E_{13}}$}
\label{Fig.: FigE14PathLS}
\end{figure}

We now pass to analyze the behaviour of the system when the exciton starts on a a high energy eigenstate
$\ket{E_{13}}$ which is mostly localized on the lumenal side (in particular on the site $b606$). The
behaviour of the populations shown in fig. \ref{Fig.: FigE14PathLS} accounts for the presence of a
bottleneck in the energy pathways \cite{Fleming09,Grondelle06}. The latter is due
on one hand to the high spatial localization of the exciton involving the site $a604$ (which is far from
the other a-lumenal sites), and on the other hand to the large energy separation with respect to the excitons
localized on the neighbouring lumenal sites $b605,b606,b607$ ($\Delta E \approx 172 cm^{-1}$ eigenvalue $8$
with respect to eigenvalue $9$ in fig. \ref{Fig. : HamSpectrum}).
Indeed, while in the first few picoseconds, the population of the b-lumenal sites decreases in favor of the
population of a-lumenal band, part of the b-lumenal population start to flow toward the b-stromal band. The
overall effect is that the a-stromal sites, and therefore the output sites, become populated with a smaller
rate than in the stromal-stromal case.

\begin{figure}
\fbox{\includegraphics[width=8cm,viewport= 1 280 640 800,clip]{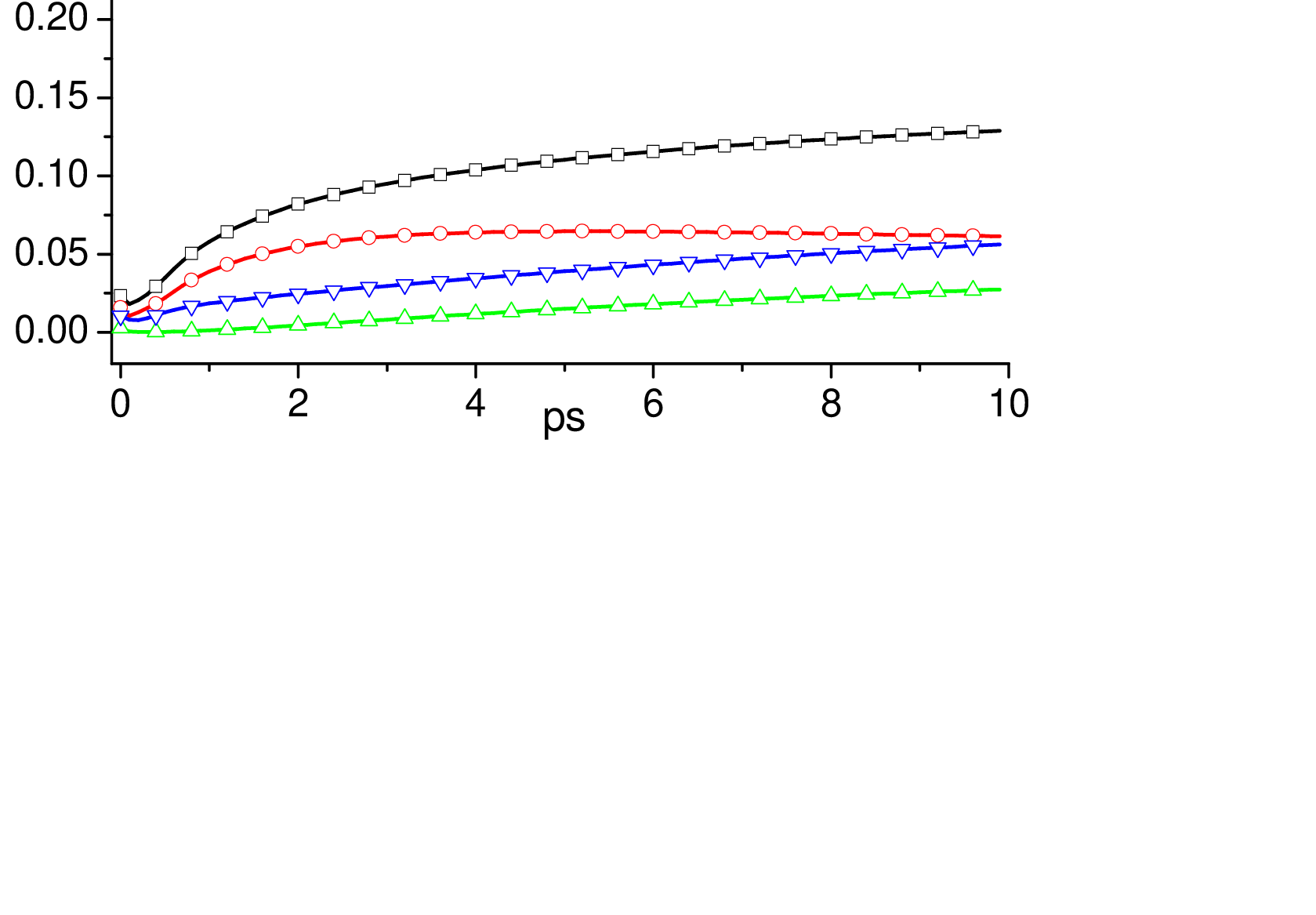}
\includegraphics[width=8cm,viewport= 1 280 640 800,clip]{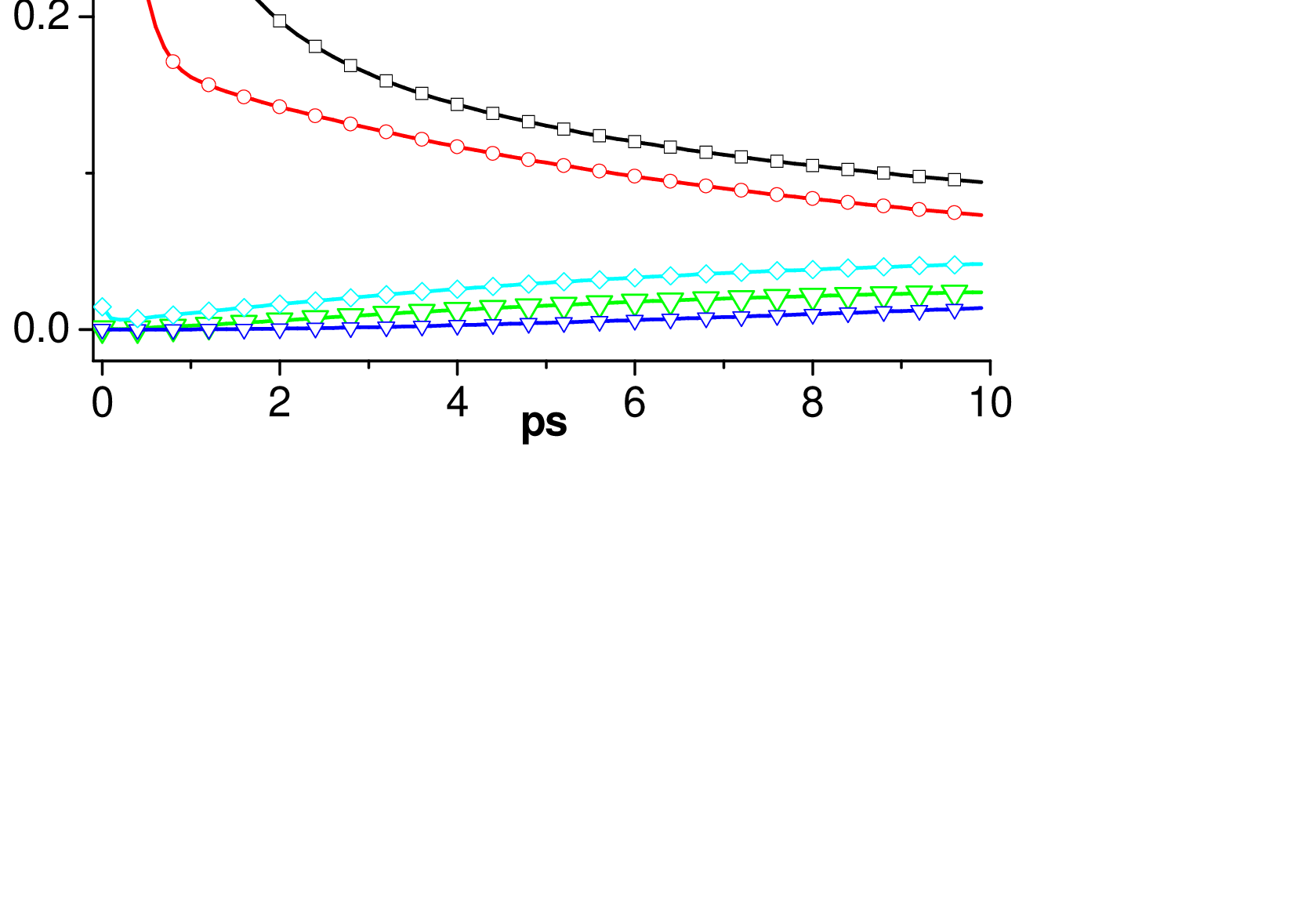}}
\caption{Lumenal-Stromal pathway: mutual information for initial state $\ket{E_{13}}$; left: stromal side;
right: lumenal side}
\label{Fig.: MIE13Tr0Deph3}
\end{figure}

The trapping effect on the lumenal side can also interpreted in terms of correlation pathways. As it is shown in the right plot
of fig. \ref{Fig.: MIE13Tr0Deph3} the b-lumenal band is initially highly correlated with the b-stromal band
(black line, right plot).

On one hand this is due to the fact that the eigenstate $\ket{E_{13}}$ is partially delocalized on the
sites $b608,b609$. On the other hand, as already mentioned, the b-lumenal to b-stromal flow, was already
pointed out  in terms of energy pathways in \cite{Grondelle06}, where the transfer time
between b-lumenal sites and b-stromal sites were estimated of the order of $2-4 ps$, and, despite the small
level of interaction ($\leq 11 cm^{-1}$, see Ham) it is consistent with the contiguity of the b-lumenal and
b-stromal sites in the monomer.

While the correlations between b-lumenal and b-stromal bands rapidly decay in the first few picoseconds,
there is not a corresponding growth of correlations on the lumenal side: the b-lumenal and a-lumenal sites
remain very poorly correlated among each other and with the rest of the a-stromal sites. At the same time
there is a clear enhancement of the correlations on the stromal side, which become rapidly greater than those
on the lumenal side, suggesting the activation of the stromal-stromal pathway.

\begin{figure}
\fbox{\includegraphics[width=8cm,viewport= 1 280 640 800,clip]{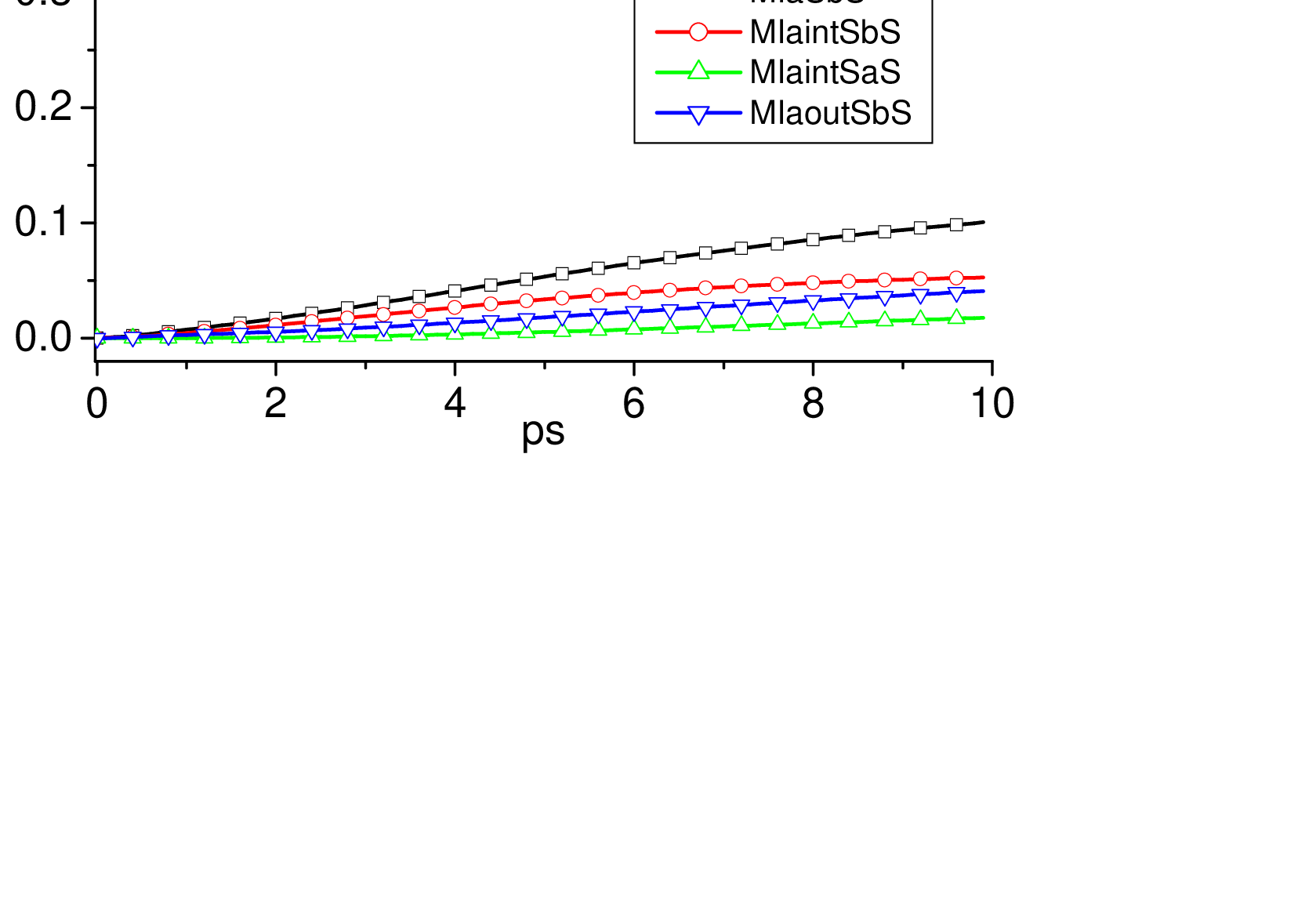}
\includegraphics[width=8cm,viewport= 1 280 640 800,clip]{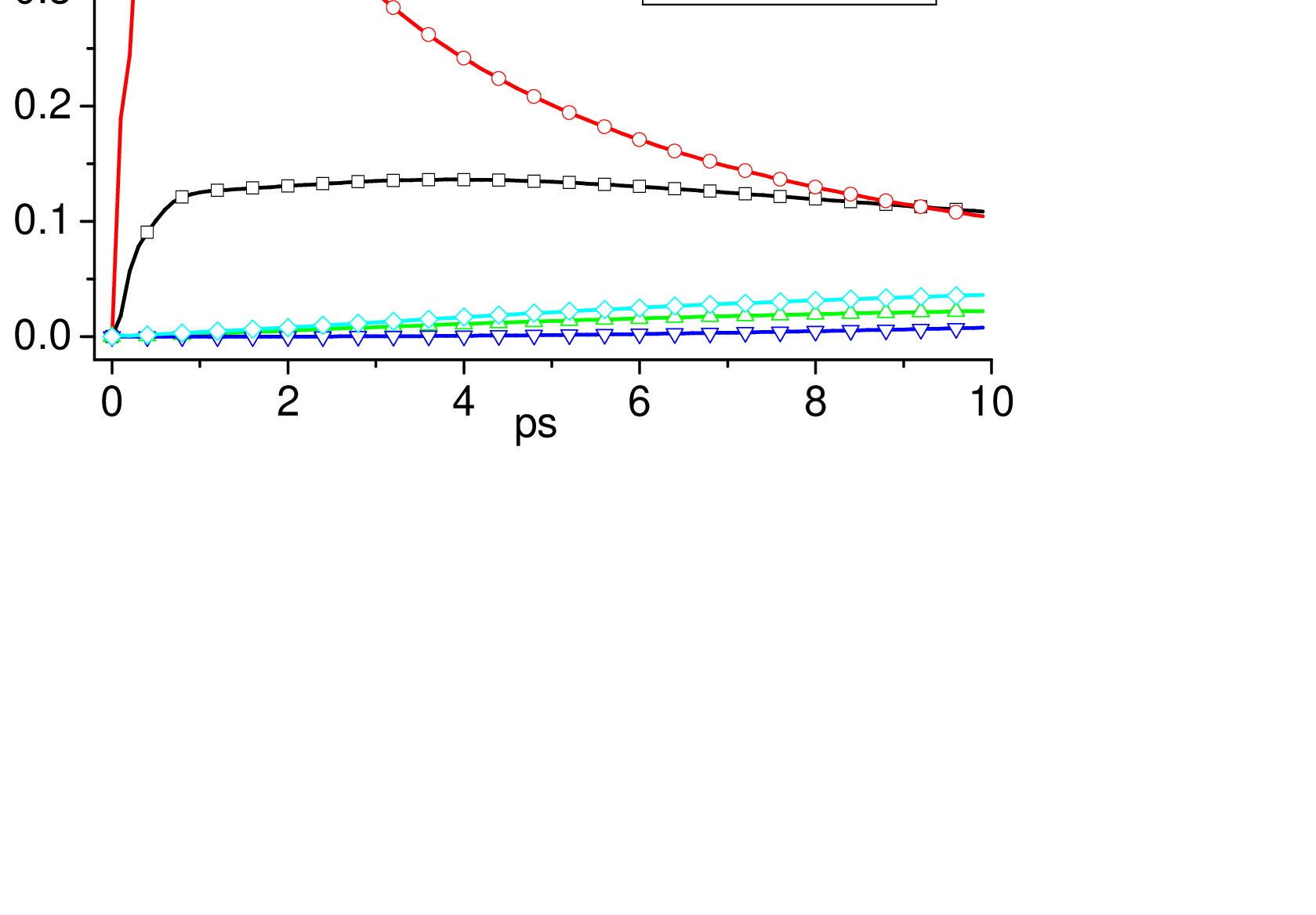}}
\caption{Lumenal-Stromal pathway: mutual information for initial state $\ket{b607}$; left: stromal side;
right: lumenal side}
\label{Fig.: MIS7Tr0Deph3}
\end{figure}

The above analysis is robust with respect to the choice of the initial state; for example in fig.
\ref{Fig.: MIS7Tr0Deph3} the same simulations have been carried out for an initial state $\ket{b607}$
fully localized on the b-lumenal site $b607$. Here the activation of the stromal-stromal pathway is
evident; in the first $1ps$ the b-lumenal subsystem gets correlated both with the sites $b605,a604$
($MIabLbL$)and with the b-stromal band ($MIbLbS$ right plot); then the correlations are mostly built on the
stromal side (left plot) while on the same time scales the correlations with the output sites on the
lumenal side are build with a slower rate.

The net effect of the presence of the above mentioned bottleneck
on the lumenal side is a substantial slowdown of the lumenal-stromal
dynamics. There are a number of possible
functional advantages for this slowdown.
One possibility is to assist in photoprotection,
the elimination of triplet exciton states
that can create harmful singlet oxygen.  This elimination takes
place primarily by the transfer of triplet excitons to
triplet carotenoid states, and has a relatively slow
timescale (a fraction of a microsecond) compared with singlet
exciton transfer \cite{Photoprotect1}.   Triplet transfer is a Dexter process,
mediated by wave function overlap, and requires the
carotenoids to be physically close to the chromophore
carrying the triplet.  X-ray crystallography studies
of LHCII suggest that the $a604$ chromophore in the lumenal
bottleneck is close enough to a lutein carotenoid to allow
triplet transfer, a process confirmed by spectroscopy
\cite{NatureSpinach, Photoprotect1}.

%(This chromophore is also
%close to a neoxanthin carotenoid which is known to play a
%significant role in photoprotection, albeit apparently not by triplet
%quenching, but instead by `plugging a hole' in the protein
%structure that allows oxygen to enter the LHCII complex
%close to the trapping region.)

A second possible function for the lumenal bottleneck
is that the bottleneck $b605$ chromophore could mediate
excitonic transport from one trimer to another \cite{NatureSpinach}.
This chromophore `sticks out' from the others in the
LHCII crystallographic structure, giving both weaker
couplings to the other chromophores within the LHCII monomer,
and potentially stronger couplings to chromophores
in neighboring triples.  Crystallographic investigations
of LHCII suggest that the $b605$ is positioned to transfer excitons to
the $b606$ and $a614$ chromophores of neighboring
trimers in LHCII aggregates \cite{NatureSpinach}.
This transfer pathway could also participate in
photoprotection via non-photochemical fluorescence quenching.

A third possible reason for the slower lumenal pathway
is that it might allow two excitons to propagate through the
LHCII complex simultaneously without quenching.
The weak coupling between the bottleneck chromophores
of the lumenal pathways and the chromophores of the
stromal pathway, together with their relative spatial
separation, could allow an exciton localized in the
lumenal pathway to wait for a stromal exciton in the
same complex to pass through the stromal trapping
states, before passing through itself.
%The slowdown is consistent with the proposed use of
%the bottleneck as a trapping mechanism for potentially harmful
%excitonic triplets via transfer to a carotenoid [ref.].

To summarize, the dynamics of the slow lumenal side and the fast stromal
side have many have a rich set of potential biological
functions.  These dynamics are in turn based on a rich quantum structure,
which the next section elucidates.

\subsubsection{Delocalization and quantum correlations}
\begin{figure}
\fbox{
\includegraphics[width=0.45\textwidth,viewport= 1 280 640 800,clip]{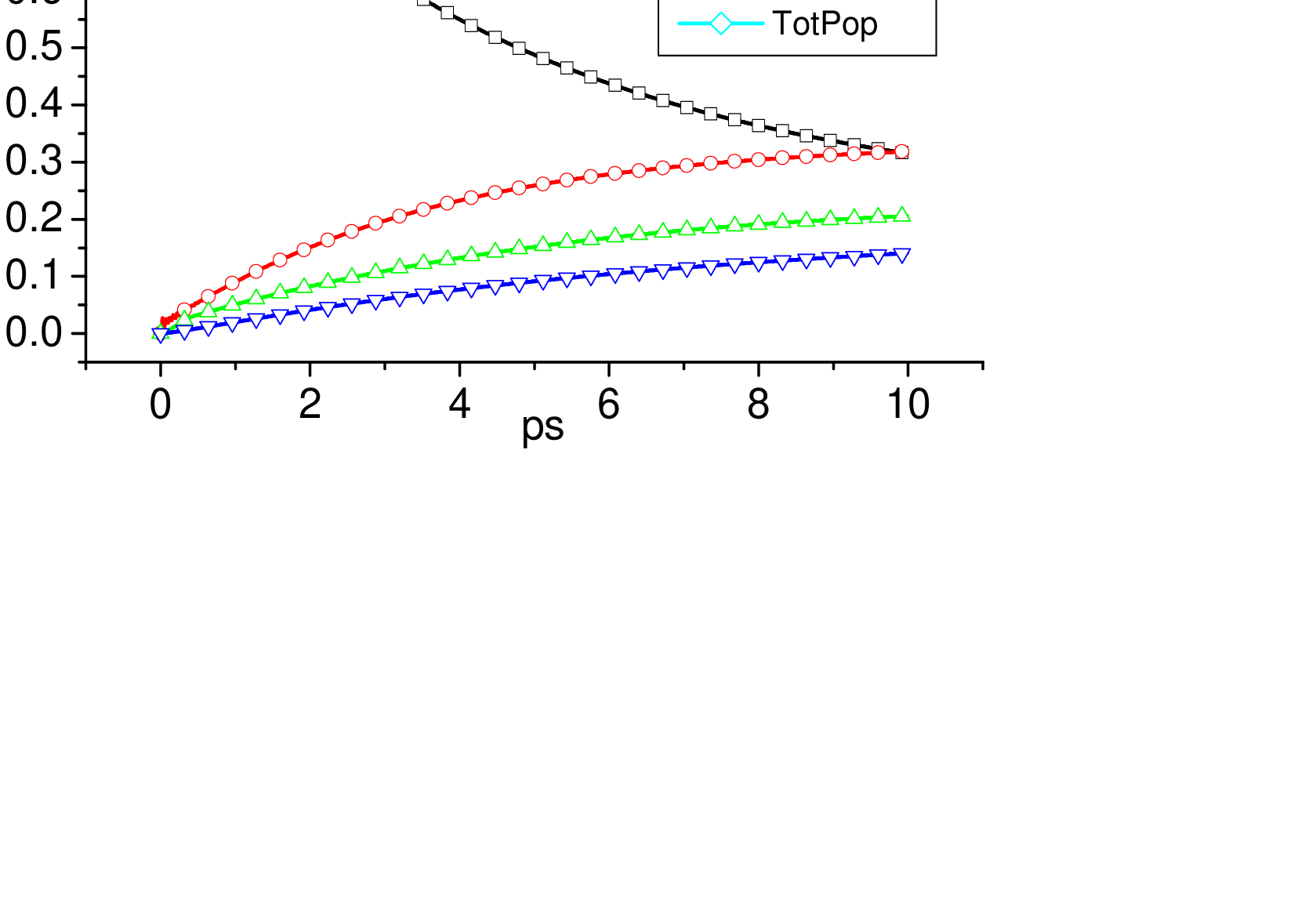}
\includegraphics[width=0.45\textwidth,viewport= 1 280 660 800,clip]{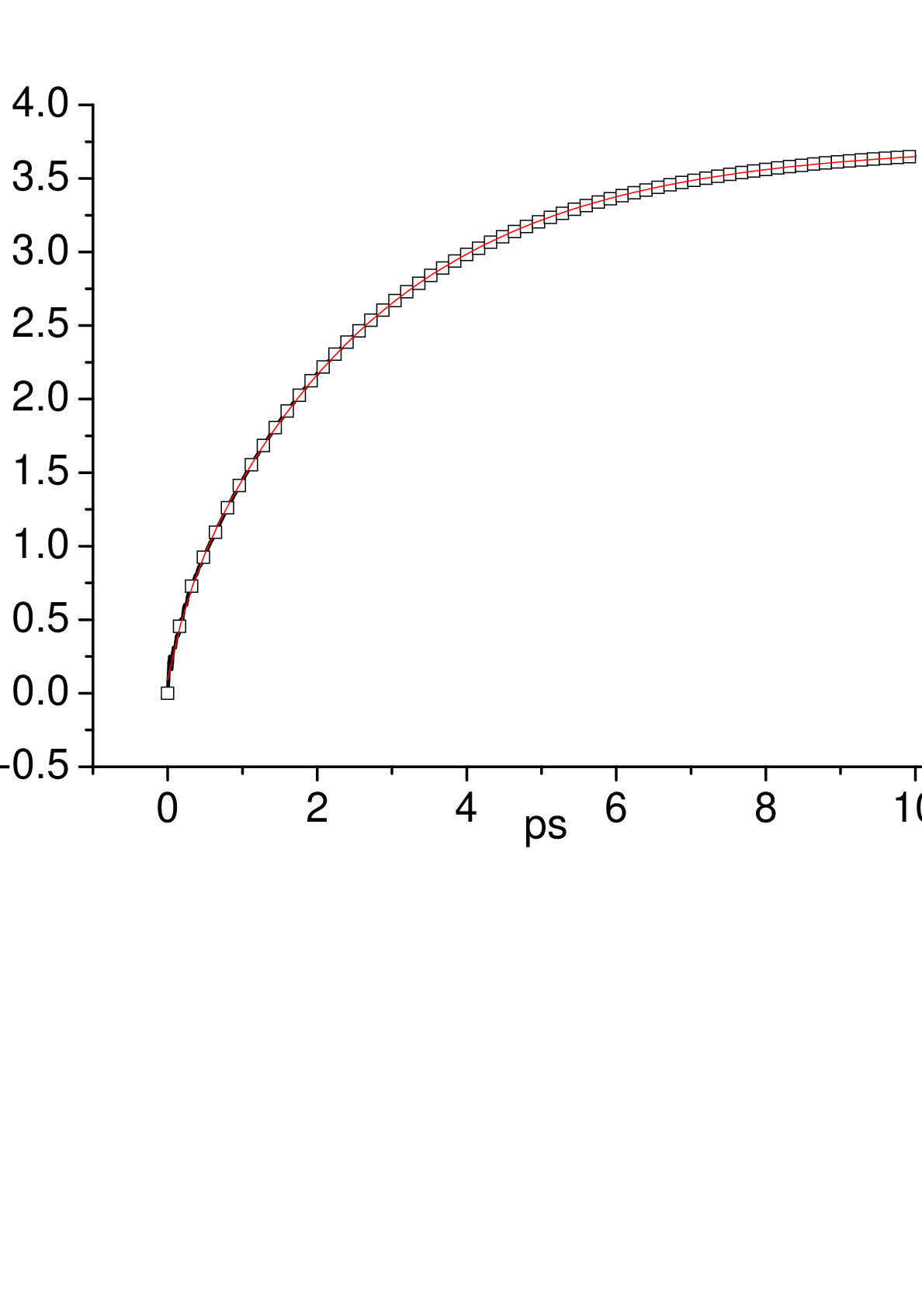}}
\caption{Initial state $\ket{b601}$ ($\approx \ket{E_{14}}$), state localized on the Chl $b601$ on the
stromal side. Populations (left) and delocalization (right) $\mathcal{D}(t)$ with fitting curve.}
\label{Fig.: DelocS1Tr0Deph3}
\end{figure}

In the following we examine the monomer dynamics from the point of view of the time scales that
characterize the delocalization of the excitons over the whole
monomeric structure and the quantum correlations between the various subsystems.
In order to estimate the delocalization time we plot the
populations of the various bands in fig. \ref{Fig.: DelocS1Tr0Deph3} (left plot),
and the previously defined measure of delocalization $\mathcal{D}(t)$ (\ref{Eq. : Delocal}).
Since we want to study the dynamics of the spreading of the correlations we first focus on the
initial state $\ket{b601}$ localized on the Chl $b601$ only. We choose this state because it is very close
to the highest energy eigenstate $\ket{E_{14}}$, which is mostly localized
on the same site but has non-negligible quantum correlations with
the rest of the system.  We want to start with a state localized
on a single chromophore, in order to study how the
correlations spread through the structure. In fig.  \ref{Fig.:
DelocS1Tr0Deph3} we plot the delocalization function $\mathcal{D}(t)$ for $\gamma_\phi= 3 ps$ ($\approx 77
K$), $\Gamma =0.001$ and no trapping. The delocalization has a very fast growth and it can be well
represented by a function $\mathcal{D}(t)= y0+A1 \exp{-t/t_1}+A2\exp{-t/t_2}$ where two time scales
$t_1\approx 250 fs$ and $t_2\approx 2.68 ps$ appear. The relevance of these time scales can be understood
by studying the quantum correlations between the relevant subsystems.

\begin{figure}
\fbox{\includegraphics[width=8cm,viewport= 1 240 640 800,clip]{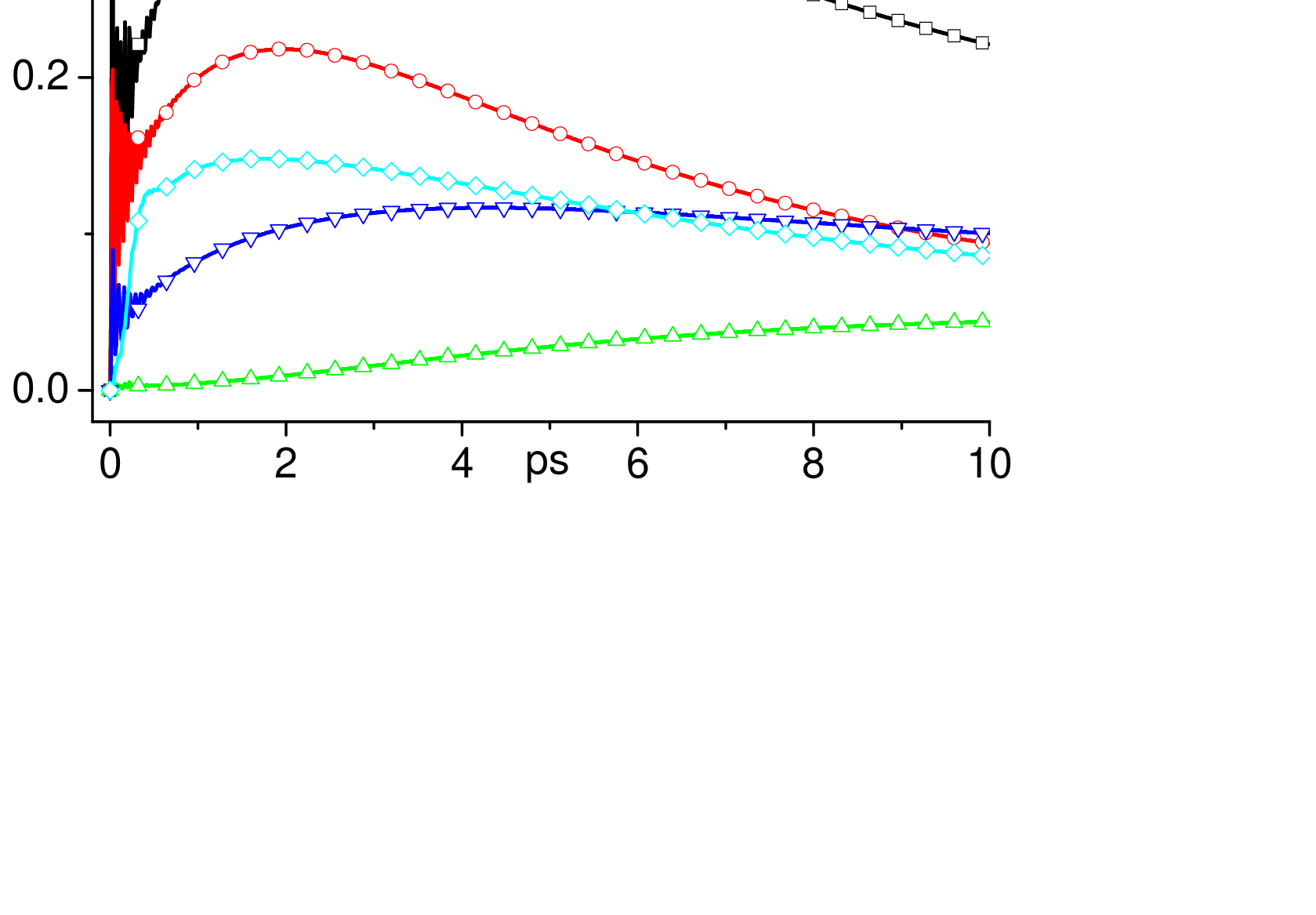},
\includegraphics[width=8cm,viewport= 1 280 640 800,clip]{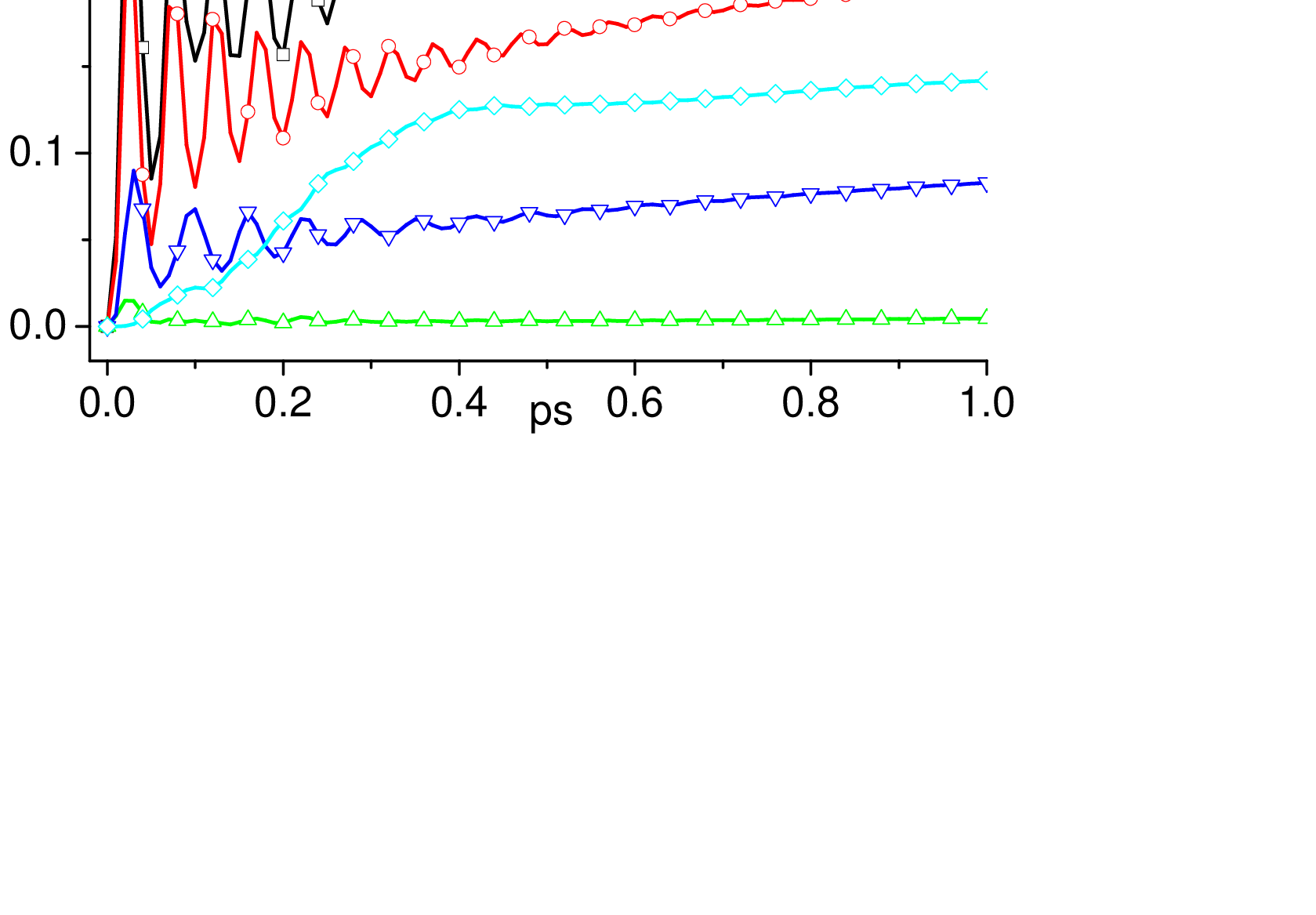}}
\caption{Initial state $\ket{b601}$ ($\approx \ket{E_{14}}$), state localized on the Chl $b601$ on the
stromal side. Mutual information between subsystems on the stromal side and between b-lumenal and b-stromal
band ($MIbLbS$); the right plot displays  the dynamics during the first $1 ps$.}
\label{Fig.: MIS1Tr0Deph3}
\end{figure}

In the left panel of fig.  \ref{Fig.: MIS1Tr0Deph3} we plot the mutual information for the relevant
subsystems over a time of $10 ps$. As it can be easily seen, the second time scale $t_2\approx 2.68 ps$ can
be correlated with the growth of the mutual information between the various subsystems which reach its maximum at a time $t$ very
close to $t_2$. In particular the correlations between the b-band and the a-band on the stromal side has a
maximum at $t\approx 2.63 ps$. In the right panel of fig. \ref{Fig.: MIS1Tr0Deph3} we show the dynamics of
correlations within the first picosecond. Here the dynamics  displays  an initial fast growth of all
correlations ($\approx 0.50 ps$)
which are characterized by an oscillating behaviour; the time scale of
these oscillations is of the order of $0.1 ps$.   These oscillations
are signatures of the high degree of quantum coherence in the
initial spreading.
The growth becomes regular within the first $400 fs\approx
1.5 t_1$; within the same period of time the correlations spread toward the lumenal side ($MIbLbS$)

\begin{figure}
\fbox{\includegraphics[width=8cm,viewport= 1 240 640 800,clip]{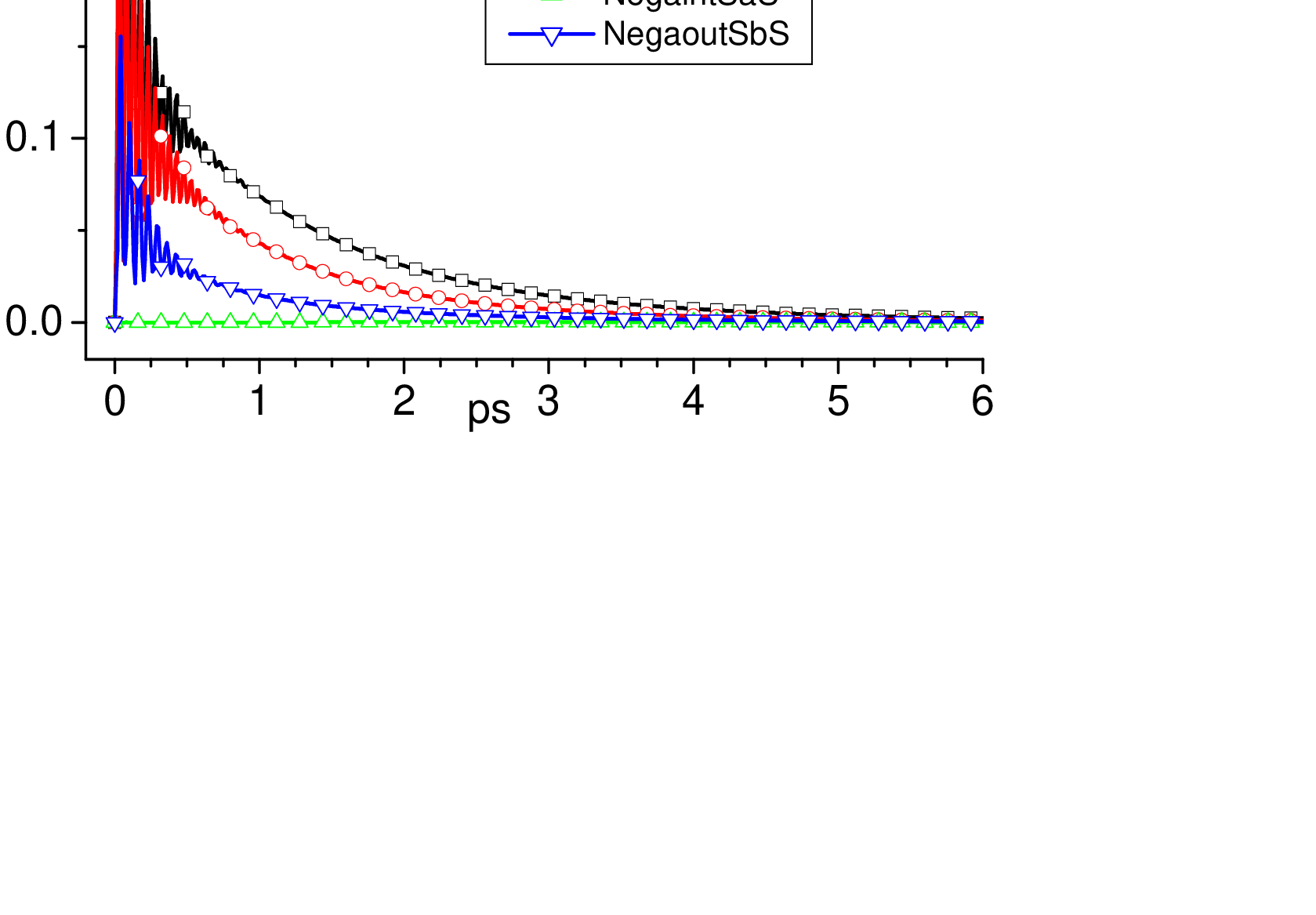},
\includegraphics[width=8cm,viewport= 1 280 640 800,clip]{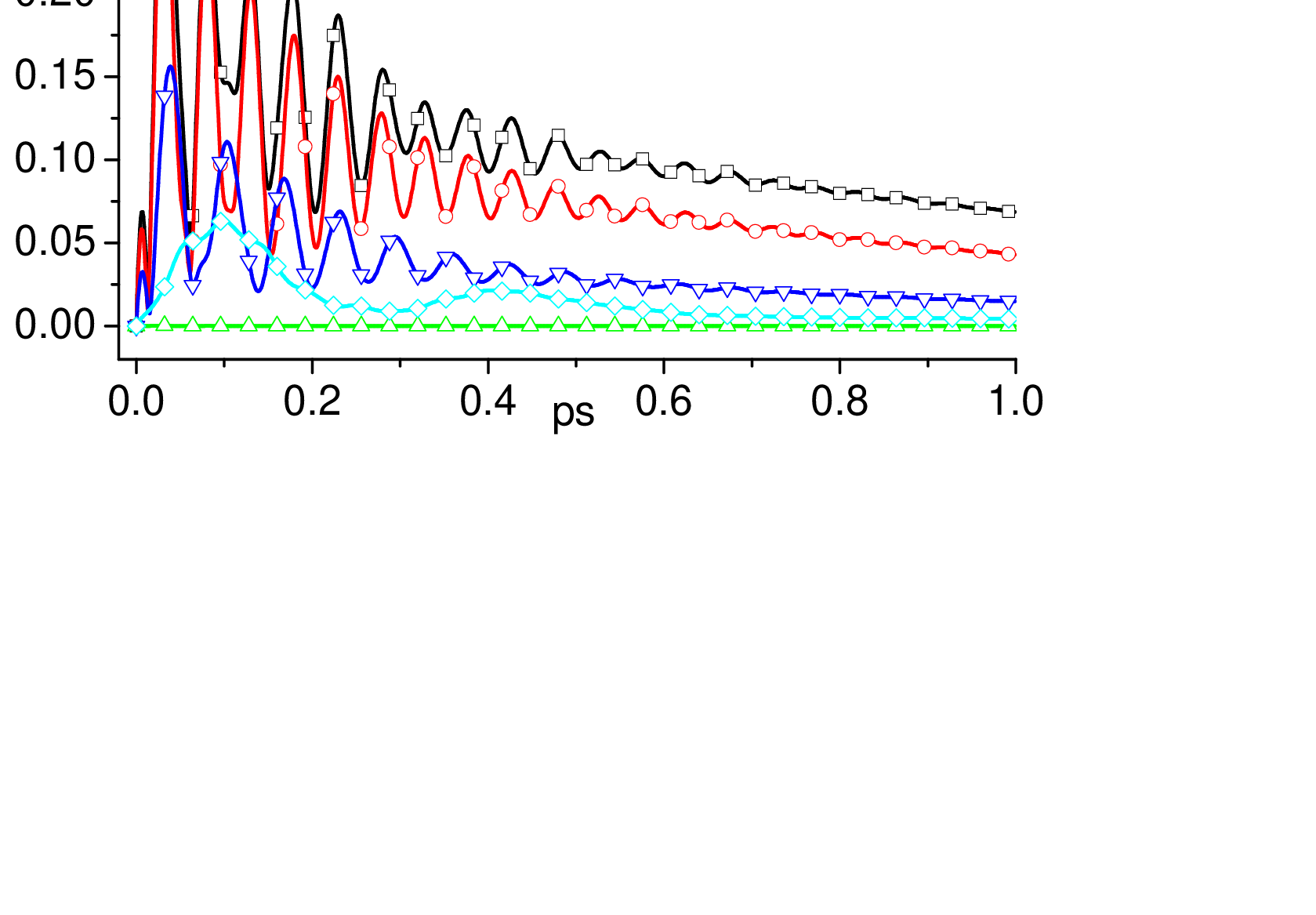}}
\caption{Initial state $\ket{b601}$ ($\approx \ket{E_{14}}$), state localized on the Chl $b601$ on the
stromal side. (Un-normalized) Negativities between subsystems on the stromal side and between b-lumenal and
b-stromal band ($NegbLbS$); the right plot displays the dynamics during the first $1 ps$.}
\label{Fig.: NegS1Tr0Deph3}
\end{figure}

\begin{figure}
\fbox{\includegraphics[width=8cm,viewport= 1 240 640 800,clip]{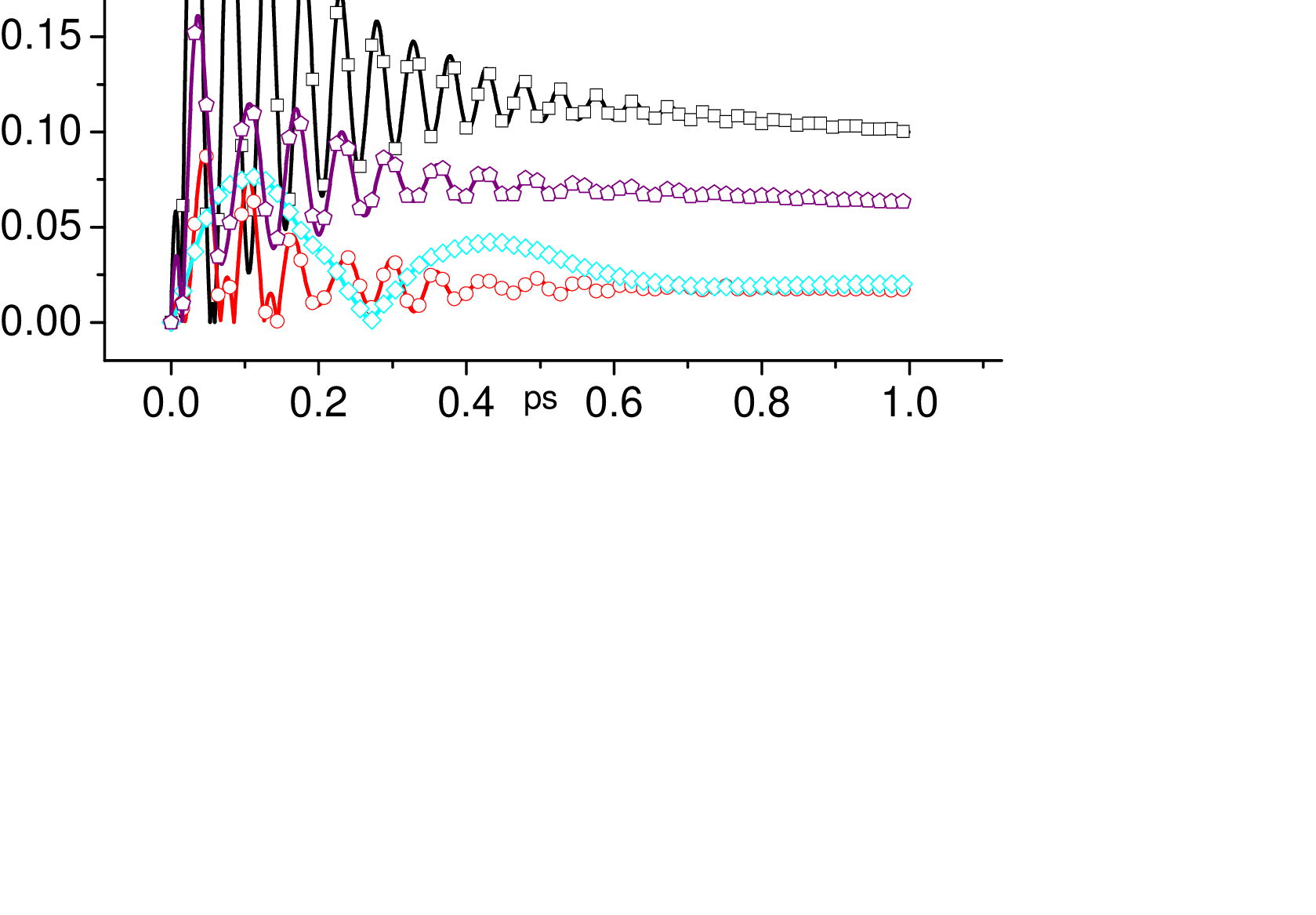}}
\caption{Initial state $\ket{b601}$ ($\approx \ket{E_{14}}$), state localized on the Chl $b601$ on the
stromal side. Relevant, i.e. non-negligible, concurrences between
site $b601$ and other sites on the
stromal ($b603$,$b11$) and lumenal ($b606$) sides;
the plot displays the dynamics during the first $1 ps$.}
\label{Fig.: ConcS1Tr0Deph3}
\end{figure}

The overall effect can therefore be described in terms of the two relevant timescales: the delocalization
takes place over the structure with an initial fast rate, and subsequently it grows toward its asymptotic
behaviour with a slower rate. The time scale of the initial rapid and
oscillating growth is consistent with the dynamics of the quantum correlations present in the system. In
fig. \ref{Fig.: NegS1Tr0Deph3} (left plot) we show the behaviour of the negativities among the relevant
subsystems over the first $10 ps$
(NegbLbS is the negativity between the lumenal and the stromal b-bands). In the first picosecond (right
plot), after a first rapid growth ($\approx 50 fs$ ) they show the same oscillatory behaviour of the mutual
information and they then decay in a smooth way after the first $\approx 400 ps$ (left plot).
This behaviour is also shown by the concurrences between sites: in fig. \ref{Fig.: ConcS1Tr0Deph3} we show
the relevant (non-negligible) concurrences between the site $b601$ and other sites.
Non-zero negativity and concurrence demonstrate the presence of
entanglement during the initial coherent spreading, similar to the
presence of entanglement in FMO \cite{Sarovar10, Caruso10}.
The quantum correlations are therefore established in the first few hundreds of femtoseconds and they give
their contribution for the first rapid growth of the delocalization of the exciton over the whole structure
(stromal and lumenal side).

%When the initial state of the simulation is an excitonic high energy eigenstate of our Hamiltonian
%($\ket{E_{13}}$ or $\ket{E_{14}}$), even if it is mostly localized on some site ($b607$ or $b601$), then
%quantum correlations are already present in the state.  These pre-existing correlations also exhibit
%oscillations with a subpicosecond time scale, dying out after a picosecond.

\subsection{Effect of the trapping on the monomer dynamics}
\begin{figure}
\fbox{\includegraphics[width=8cm,viewport= 1 240 640 800,clip]{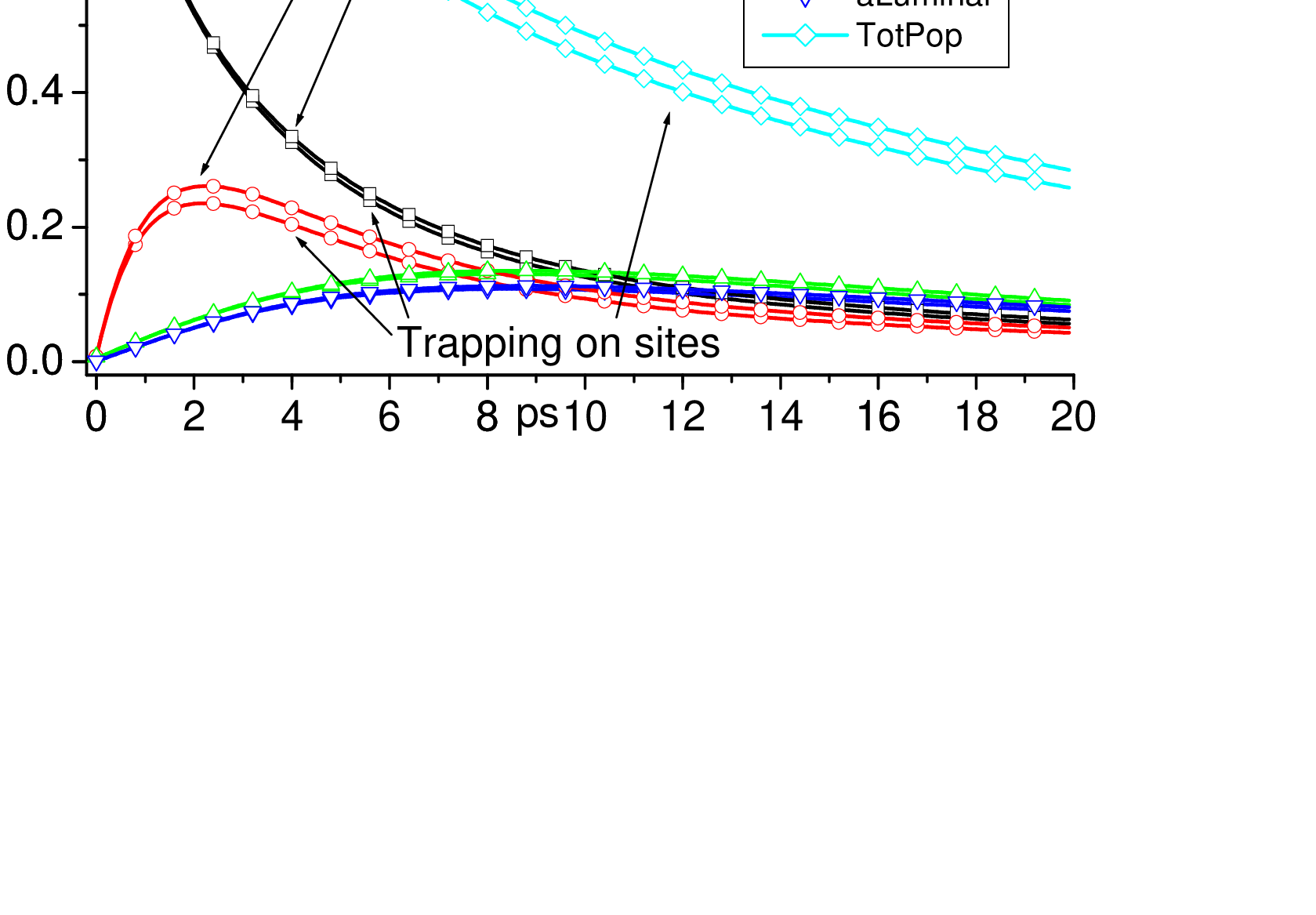}
\includegraphics[width=8cm,viewport= 1 240 640 800,clip]{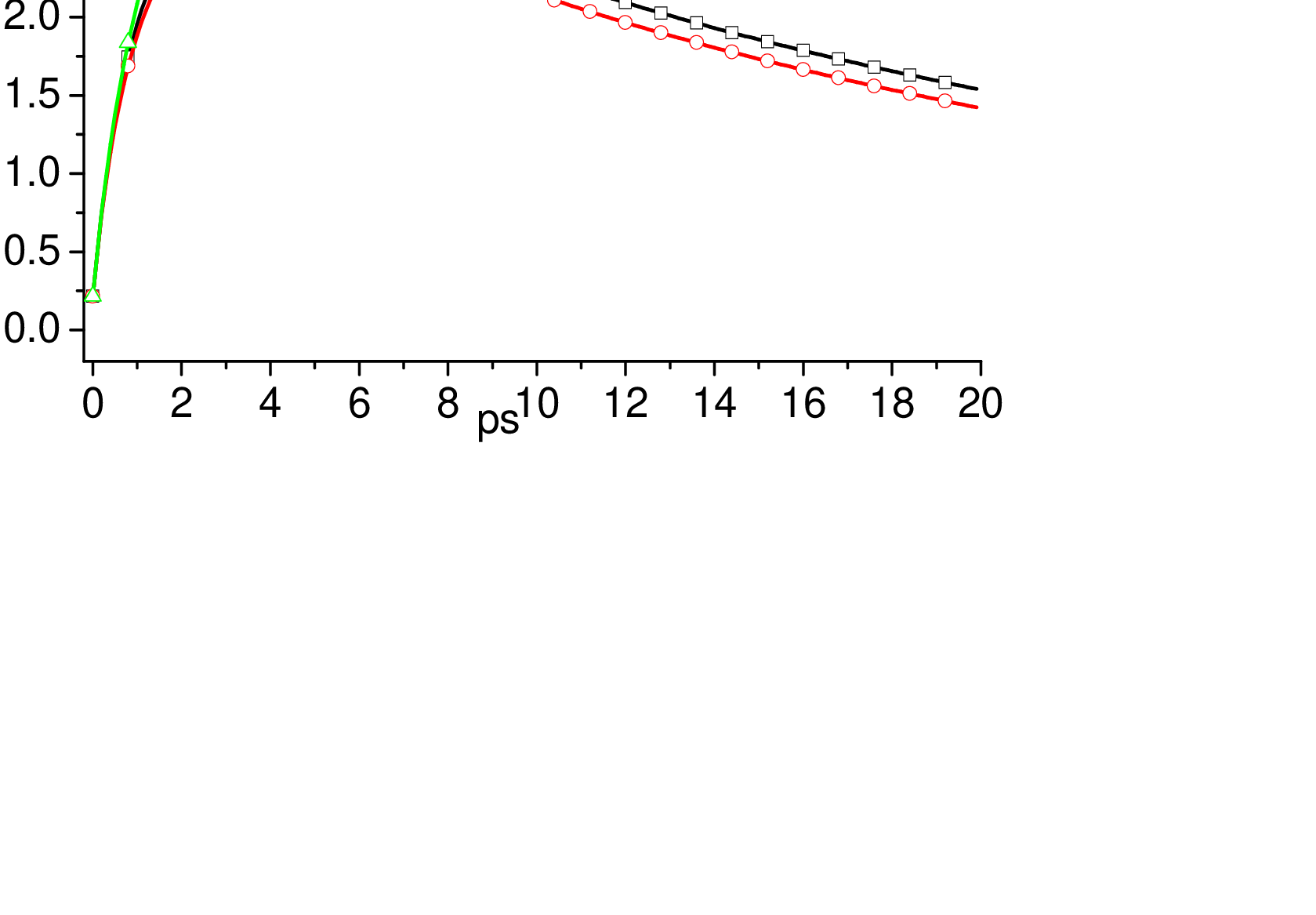}}
\caption{Initial state $\ket{E_{14}}$. Dephasing $\gamma_\phi= 12 ps^{-1}$; trapping $k_{trap}=1 ps^{-1}$;
recombination $\Gamma=0.001 ps^{-1}$. Left:  populations of the different bands with exciton/site base
trapping; right: delocalization functional $\mathcal{D}(t)$ with exciton/site base trapping and no
trapping}
\label{Fig.: monoE14D12EvsSTr1}
\end{figure}

In order to see how the trapping modifies the monomer dynamics we focus on the stromal-stromal dynamics.
The initial exciton state is $\ket{E_{14}}$ and we choose to fix the value of the dephasing rate to
$\gamma_\phi=12 ps^{-1}$ (that approximately corresponds to the ambient temperature if one uses \ref{Eq. :
gamma_phi}).
In fig. \ref{Fig.: monoE14D12EvsSTr1}, the left plot displays the populations of the various subsystems.
The comparison of the populations dynamics with the site based trapping mechanism and the exciton based
mechanism shows that the first one is obviously more efficient in reducing the total population and the
population of the various band (in particular the a-lumenal one).  As
noted above, this difference arises in our model because
site-based trapping operates on three sites, while the
exciton based mechanism acts only when the two lowest eigenstates get
populated.

In the right plot of fig. \ref{Fig.: monoE14D12EvsSTr1}  we show the delocalization functional for no
trapping and for site/exciton based trapping. The presence of the trapping mechanism starts to be relevant
already after the first picosecond, i.e. before the full delocalization of the exciton has occurred. Indeed
for $\gamma_\phi=12 ps^{-1}$ the typical time scales of the delocalization without trapping are $t_1\approx
400 fs$ and $t_2\approx 1.5 ps$. This means that, in presence of trapping, the delocalization due to the
initial dynamics of the quantum correlations has a fundamental role in the energy transfer process.

\subsection{The monomer as wire}
An interesting problem is to determine the typical time scales that govern the delocalization of an exciton
initially localized on the output sites $a610,a611,a612$ of the monomer. These quantities become relevant when one describes
the behaviour of the dimer and the trimer when they act as quantum wires.
Indeed, the LHCII complex can in principle be activated by other
neighbouring LHCII complexes and this should happen when the output sites
of a pair of LHCII are sufficiently close.  In this process,
the $a610$, $a611$, and $a612$ sites on an LHCII monomer accept
an exciton from the same sites on a neighboring trimer.
The exciton then spreads first through the monomer, and
then throughout the three LHCII units of the trimer.
When it reaches another set of output sites, the exciton
can be transferred to another trimer, and the process repeats.

\begin{figure}
\fbox{\includegraphics[width=8cm,viewport= 1 280 640 800,clip]{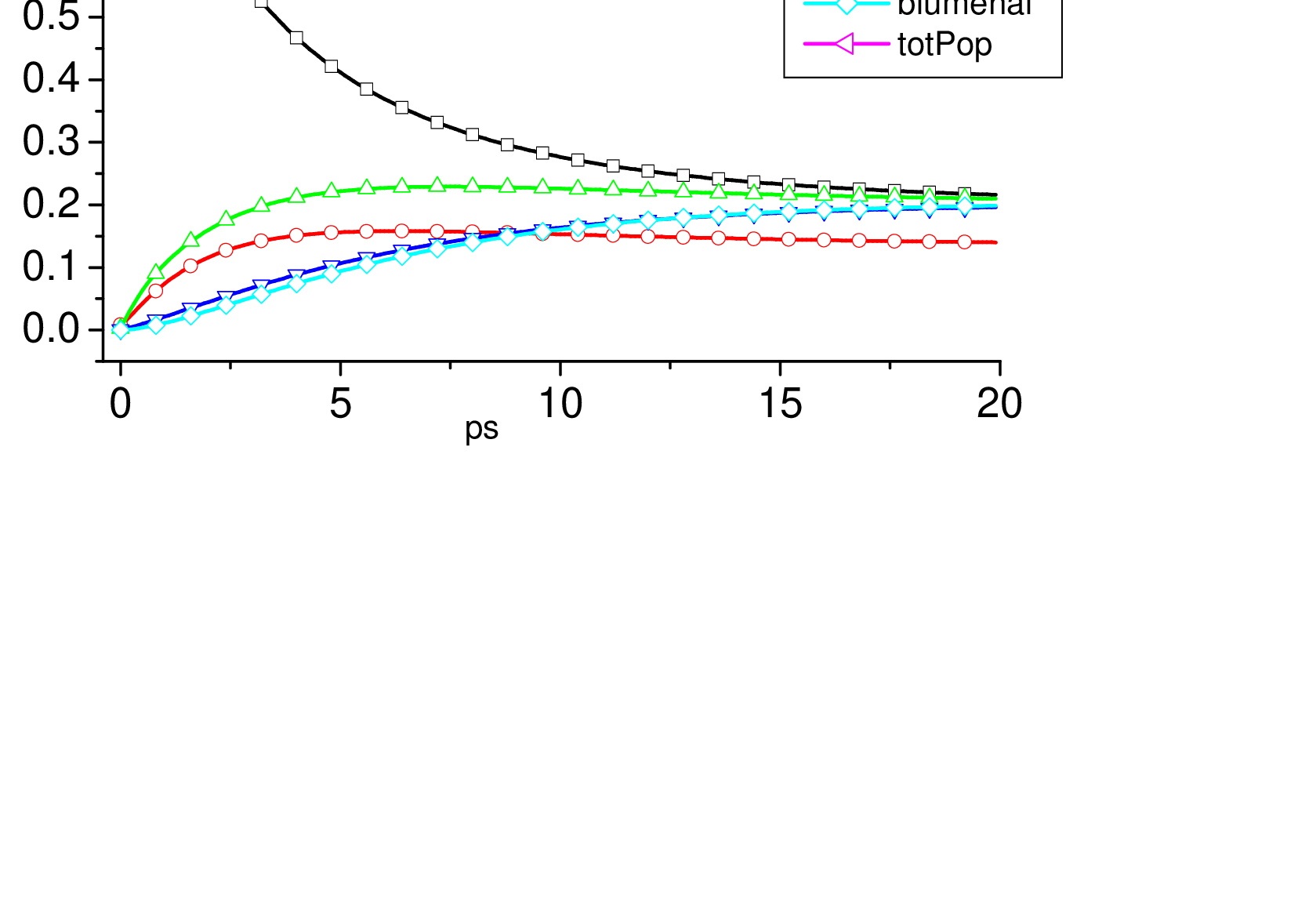}
\includegraphics[width=8cm,viewport= 1 280 660 800,clip]{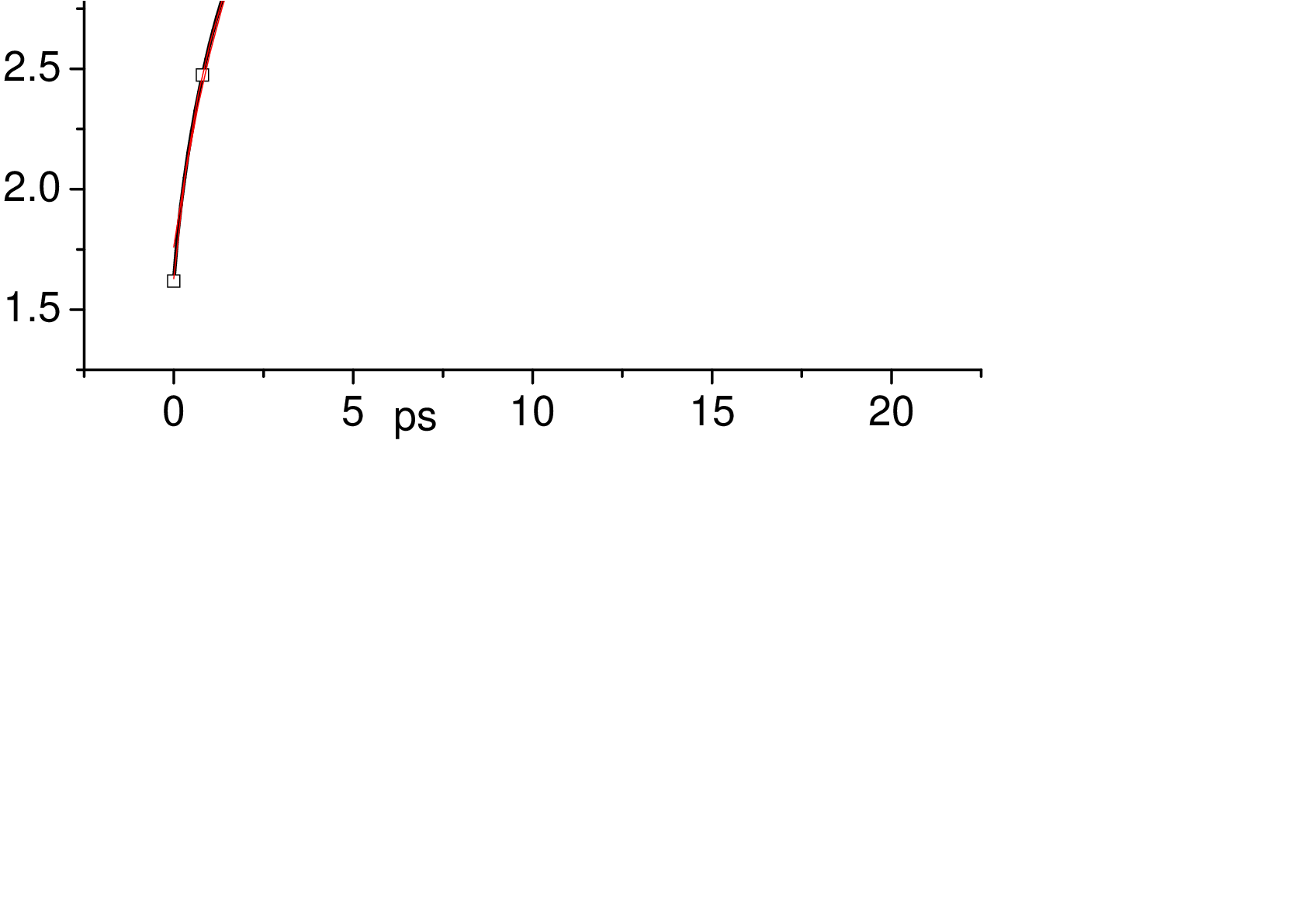}}
\caption{Initial state $\ket{E_1}$, ground state of the monomer localized over the output sites.
Populations of the different subsystems and delocalization functional $\mathcal{D}(t)$}
\label{Fig. : PopmonoE1D12T20}
\end{figure}

Accordingly, we now analyze the interaction between donors $a610,a611,a612$ on
one monomer within the trimer, and ``acceptor" $a610,a611,a612$
sites on a second monomer within the trimer.
The acceptor LHCII sites behave
as a sink with a trapping rate that depends on how fast the
exciton localized on those ``acceptor" sites diffuses to a neighboring
LHCII trimer or to some other part of the overall photocomplex. In order to
study this diffusion process, we first focus on a single monomer, initialized
in the monomer Hamiltonian ground state
$\ket{E_{1}}$ (localized on $a610,a611,a612$).
In order to estimate the
delocalization time we plot the populations of the various bands in fig. \ref{Fig. : PopmonoE1D12T20} (left
plot). We use the previously introduced delocalization
measure $\mathcal{D}(t)$ which is plotted in the
right part of fig. \ref{Fig. : PopmonoE1D12T20}.
The simulations employ a fixed value of the dephasing
rate $\gamma_\phi=12 ps^{-1}$ that approximately corresponds
to ambient temperature.
The population of the acceptor sites decreases and becomes of the same order of the other populations in
about $10 ps$. If we take $\mathcal{D}(t)= y0+A1 \exp{t/t_1}+A2\exp{t/t_2}$ to estimate the delocalization
process over the whole monomer ($\mathcal{D}(t)$ is a site based measure of delocalization) we see that the
relevant time scale $t_2\approx 2 ps$ (while for the initial fast growth we have $t_1\approx 350 fs$).

The value of $t_2$ is of the same order of the inverse of the trapping rate which is characteristic of the
FMO $k_{trap}^{-1}=1 ps$ that we used for the monomer
in the previous section, and that we will use for the dimer, the
trimer and the tetramer in the following section.

%It would anyway be preferable to estimate this trapping rates by
%simulating the dynamics of the dimer (trimer) when its initial state is localized on the output sites of
%one monomer and with no trapping on the other monomers. This would probably give (at fixed dephasing
%rate) a better estimation of how the dimer (trimer) behaves as a sink. We should expect smaller
%delocalization time scales, since in the case of the dimer (trimer), thanks to the intra-monomeric
%couplings, there is much more "room" for the delocalization process.(TO BE DONE)

\section{Efficiency: a comparison between clusters of monomers}

In this section we examine the efficiency and typical transfer time  of the single monomer and of groups of 2, 3, and 4 monomers. The clusters
are build by connecting monomeric subunits via a site-site interaction:
as pointed out in \cite{Fleming09} there is evidence of a relatively strong coupling in the cluster $\{b601,b608,b609\}$,
where $b601$ belongs to one monomer and $\{b608,b609\}$ to another one. In \cite{NatureSpinach}, where the LHCII of the spinach is analyzed,
the link between the $b601$ and the $b609$ sites is estimated to
be $42 \; cm^{-1}$. We therefore write the overall Hamiltonian of
the complex, say the trimer, as:
\be H_{trimer}=\ensuremath\sum_{i=1,2,3} H^i_{monomer}+H_{12}^{int}+H_{23}^{int}+H_{31}^{int},
\label{Eq:. Htrimer} \ee
where the terms $H_{ij}^{int}=V_{b601_{i},b609_{j}}(\ketbra{b601_{i}}{b609_{j}}+\ketbra{b609_{j}}
{b601_{i}}$
account for the Coulomb interaction between the Chl $b601$ on the
$i-th$ monomer and the Chl $b609$ on the $j-th$ monomer. $V_{b601_{i},b609_{j}}$
is chosen to be equal to $42cm^{-1}$ \cite{Fleming09}.
We will also consider the case when the coupling $b601-b608$ is present
and equal to $42cm^{-1}$.

The recombination parameter is fixed at $\Gamma=0.001ps^{-1}$, which
again is the value used for the FMO in \cite{SethFMO2}. Trapping is supposed to be similar to
the monomeric case; the output sites are now the group of sites $a610,a611,612$
on each monomer, and the trapping Hamiltonian is $ \sum_{i}k_{trap}(\ketbra{a610_{i}}
{a610_{i}}+\ketbra{a611_{i}}{a611_{i}}+\ketbra{a612_{i}}{a612_{i}})$,
where $i$ labels the monomers, with $k_{trap}=1ps^{-1}$.

The main result of these simulations is that there is {\it a clear evidence for
a dephasing-assisted mechanism that enhances the transport efficiency} of the systems.   This mechanism
is an example of environmentally assisted quantum transport (ENAQT) \cite{SethFMO1,SethFMO2,Plenio08}.
The result is independent of the structure analyzed.
Efficiencies and the average transfer times have their optimal values in correspondence of a dephasing rate
$\gamma_\phi\geq 10 ps^{-1}$, i.e. the rate corresponding to ambient temperature.

\subsection{The antenna configuration}

 We first describe the efficiency of the various complexes when they act as antennae.
We start by focusing on the relevant figures of merit for the single monomeric unit. In fig. \ref{Fig.:
E14vsE13_eta&tau} we show the efficiency (left) and the average transfer time (right) for two different
initial states: $\ket{E_{14}}$, mostly localized on the stromal side and $\ket{E_{13}}$, mostly localized
on the lumenal side. The plots show that the differences highlighted by our analysis of the dynamics in the
previous section have relevant effects also in terms of the transport efficiency of the monomer.
The stromal-stromal pathway that starts with $\ket{E_{14}}$ results in general in a better efficiency and a
smaller $\tau$ over the whole range of dephasing rates. In particular,
at $\gamma_\phi\geq 10$ $\tau$ is $\approx 1.5$ times smaller than the corresponding lumenal-stromal value.
The mismatch in the characteristic time of the two pathways arises from the bottleneck present in the lumenal pathway.

%Two excitons could be created
%on the same monomer and, while the first exiton uses the stromal-stromal pathway
%to rich the reaction center, the second exitons would travel slower on the lumenal-stromal
%pathway allowing for the reaction center to be ready again to use it.

\begin{figure}[h!]
\fbox{\includegraphics[width=0.45\textwidth,clip]{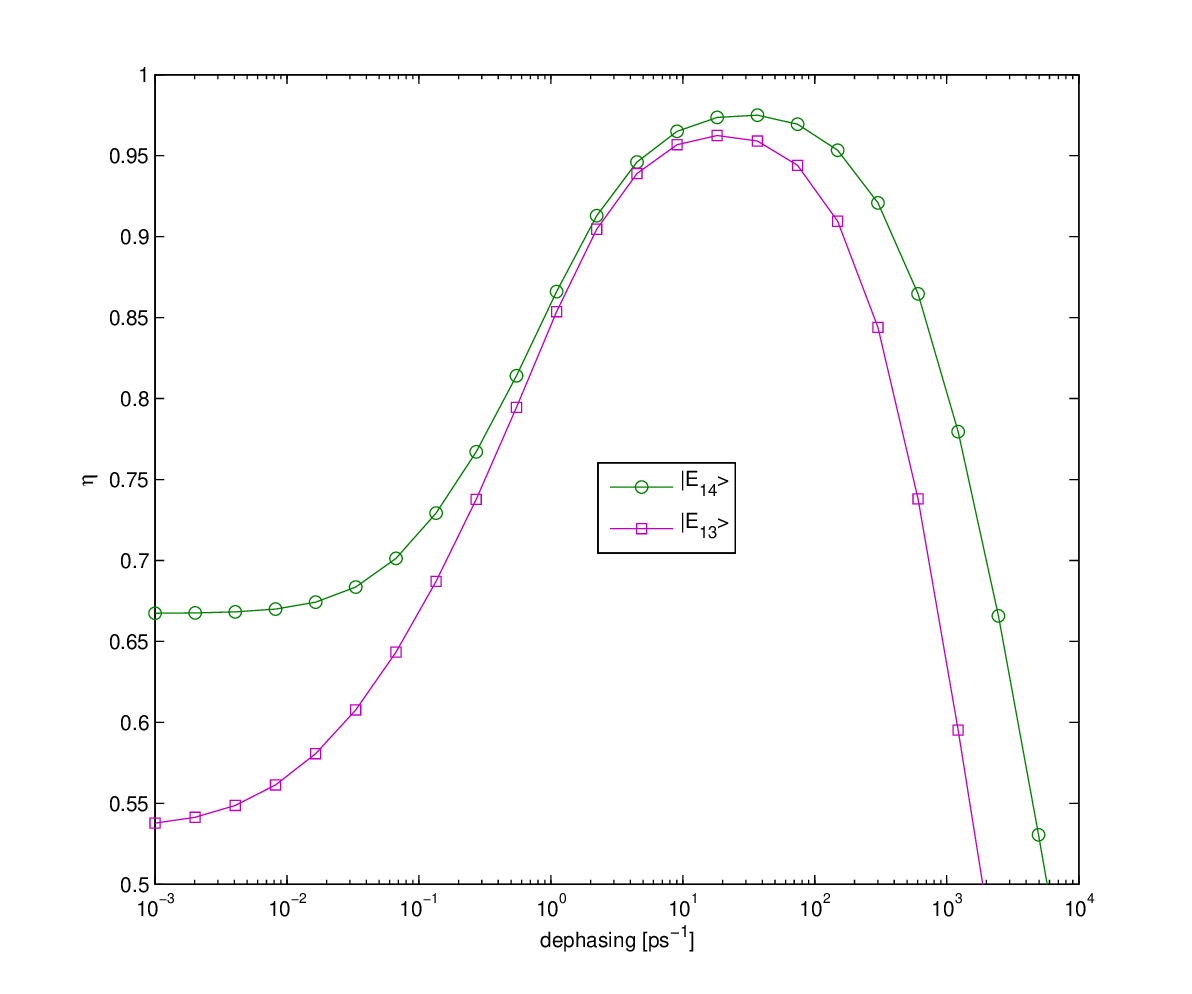} \includegraphics[width=0.45\textwidth,clip]{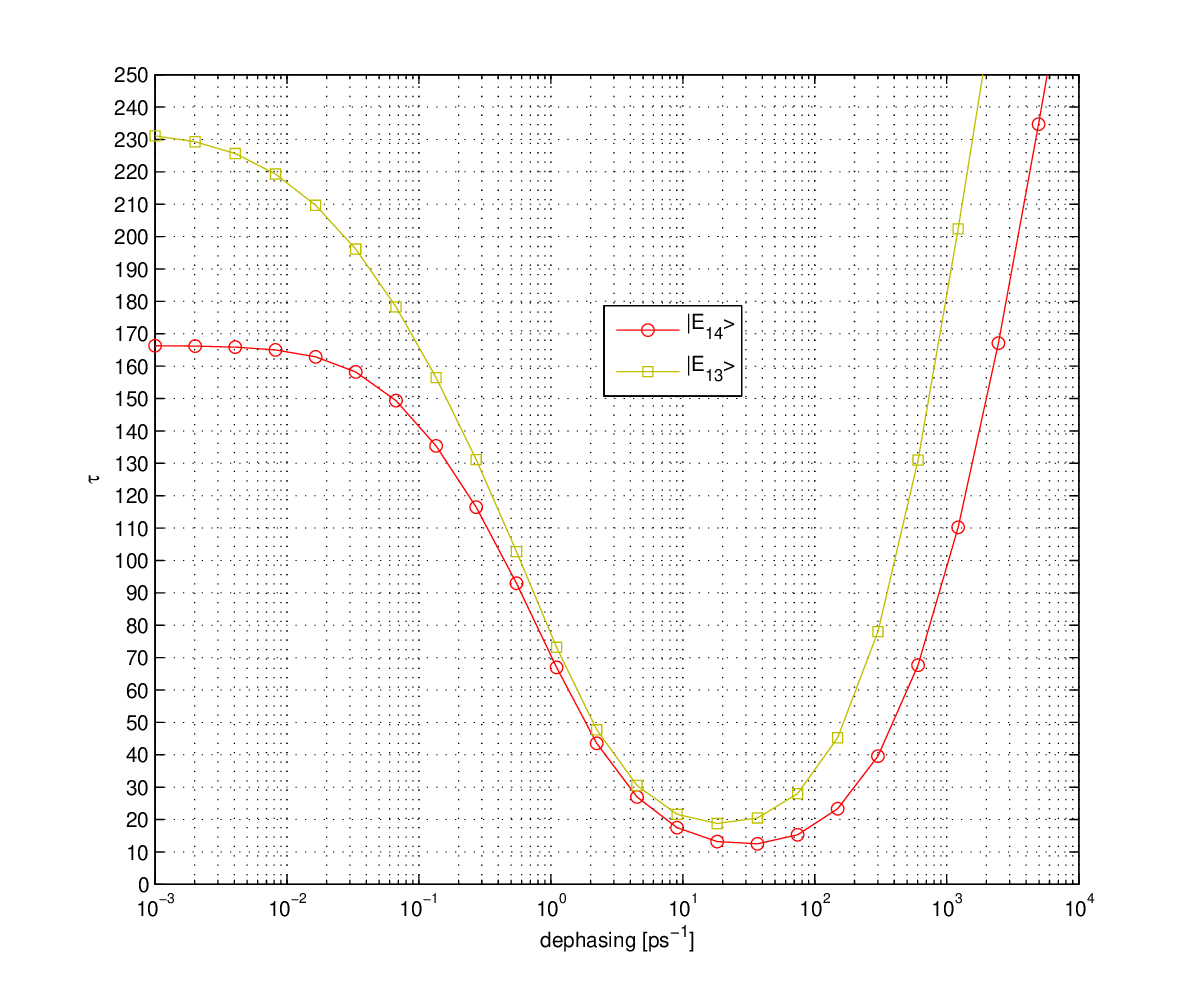}}
\caption{Efficiency $\eta(\gamma_\phi)$ (left) and Average transfer time $\tau(\gamma_\phi)$ (right) for
the monomer with different initial states: $\ket{E_{13}}$ (mostly localized on the lumenal side) and
$\ket{E_{14}}$ (mostly localized on the stromal side).}
 \label{Fig.: E14vsE13_eta&tau}
\end{figure}

\begin{figure}[h!]
\fbox{\includegraphics[width=0.45\textwidth,clip]{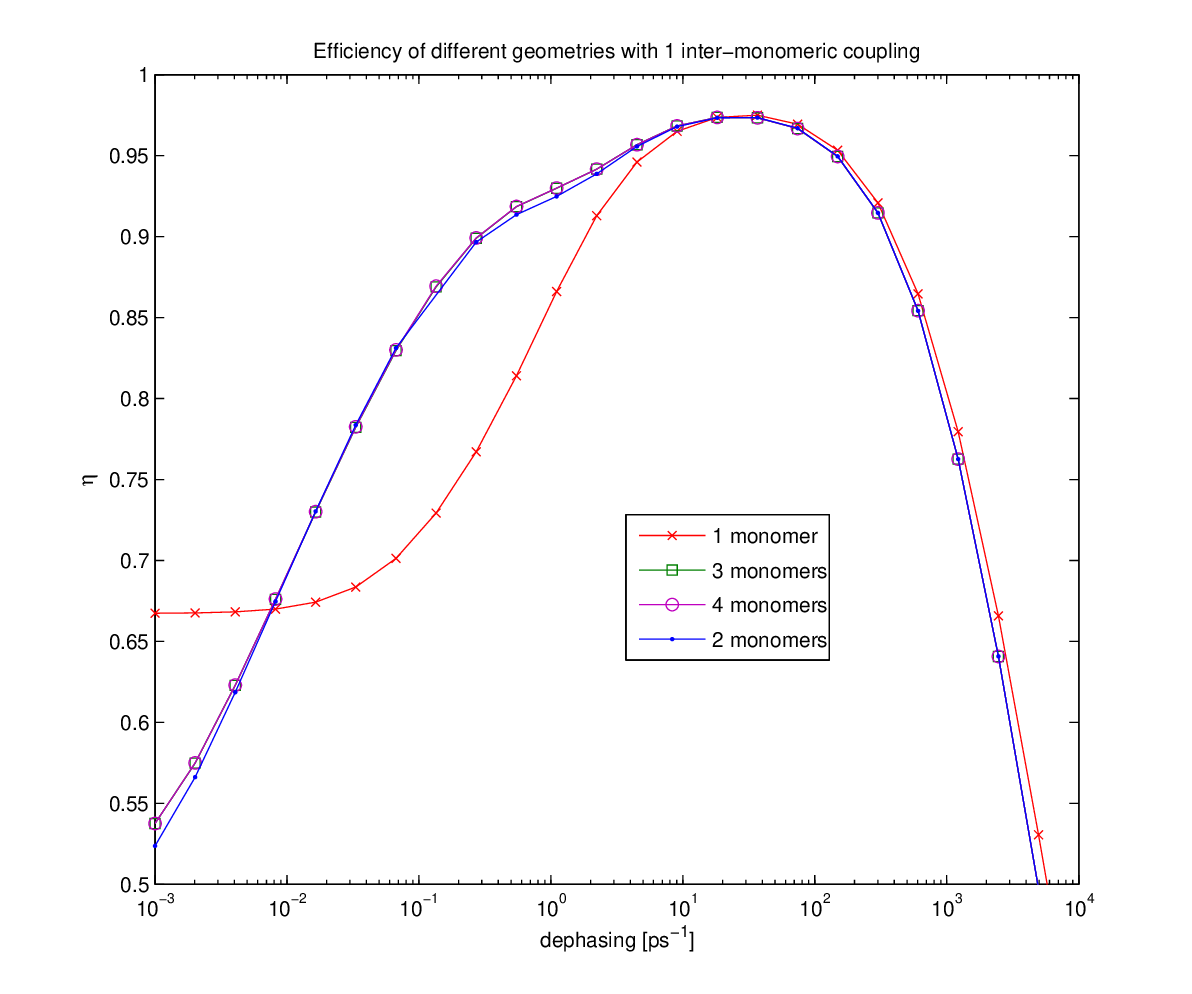}
\includegraphics[width=0.45\textwidth,clip]{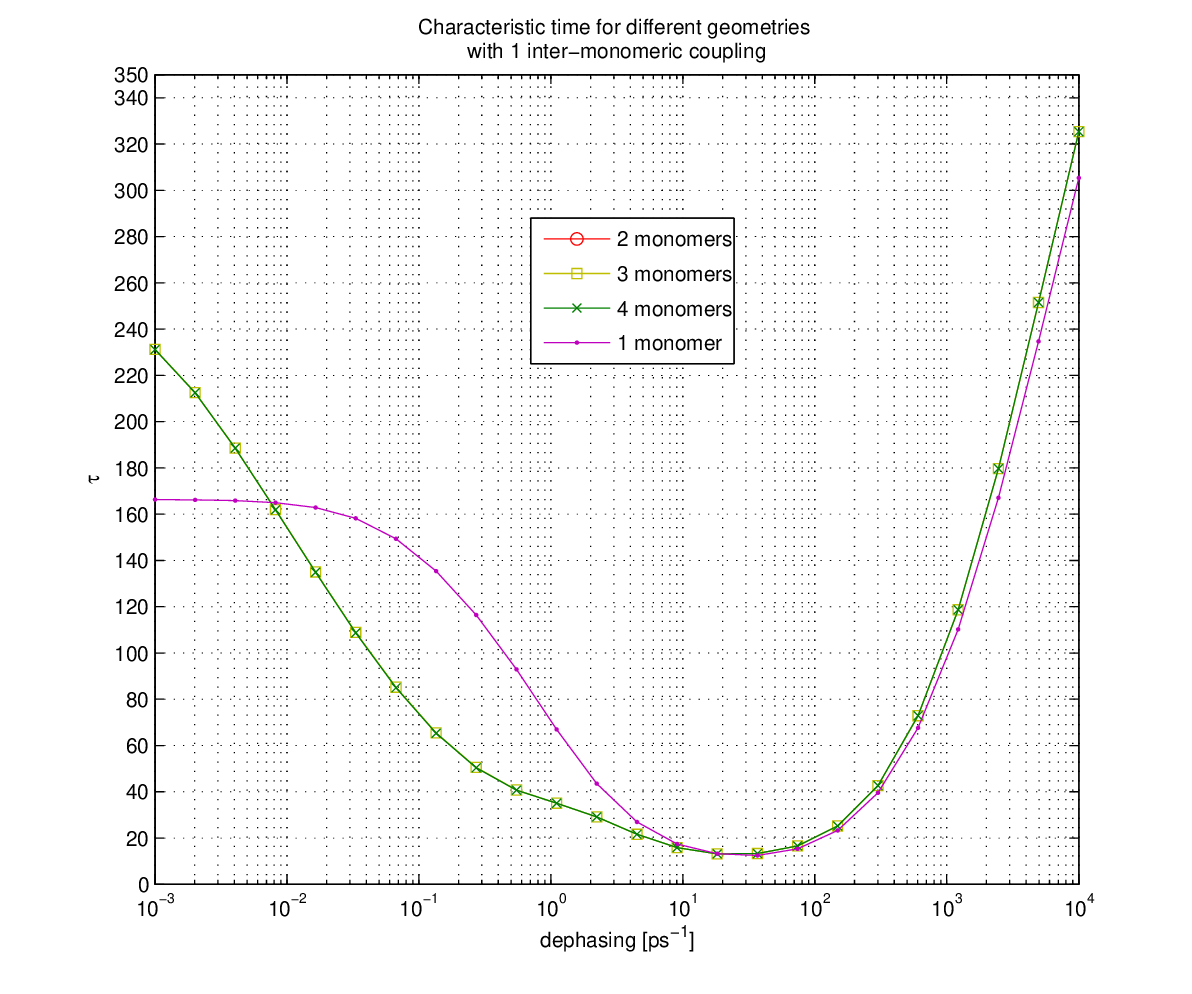}}
\caption{Efficiency $\eta$ (left) and average transfer time $\tau$ ($ps$) (right)for the different topologies
 (monomer, dimer, trimer and complex with 4 monomers)
with {\bf single} intra-monomeric coupling for various values of the dephasing rate $\gamma_\phi$. Initial
state: highest energy eigenstate of the given structure.}
 \label{Fig.: donorEtaTauC19}
\end{figure}

We now describe in detail the results of our simulations for more complex structures.
We first examine the case in which there is an active sink attached to {\em each of the monomers} of a given
structure. The initial state used for the various simulations is the
highest excited state of the
Hamiltonian, which is
basically a copy over different monomers of the eigenstate $\ket{E_{14}}$ of one single
monomer: it is thus mainly localized on the b-stromal Chls of each structure.
In Fig. \ref{Fig.: donorEtaTauC19} (left) we compare the efficiency of
the energy transport
for different geometries in presence of only one inter-monomeric coupling.
While the dephasing assisted mechanism is always present, we see that
for physically relevant values of the dephasing parameter the change of geometry
does not provide significant modifications in the efficiency.
Fig. \ref{Fig.: donorEtaTauC19} (right) shows the result of the simulations for
$ \tau $, the characteristic transfer time. The value corresponding to
the highest efficiency in the transport is around 10-15 ps.  This
transport time is consistent with that observed for the FMO complex,
taking into account the fact that each monomer subunit of the
LHCII has twice the number of chromophores of FMO.

\begin{figure}[h!]
\includegraphics[width=0.45\textwidth,clip]{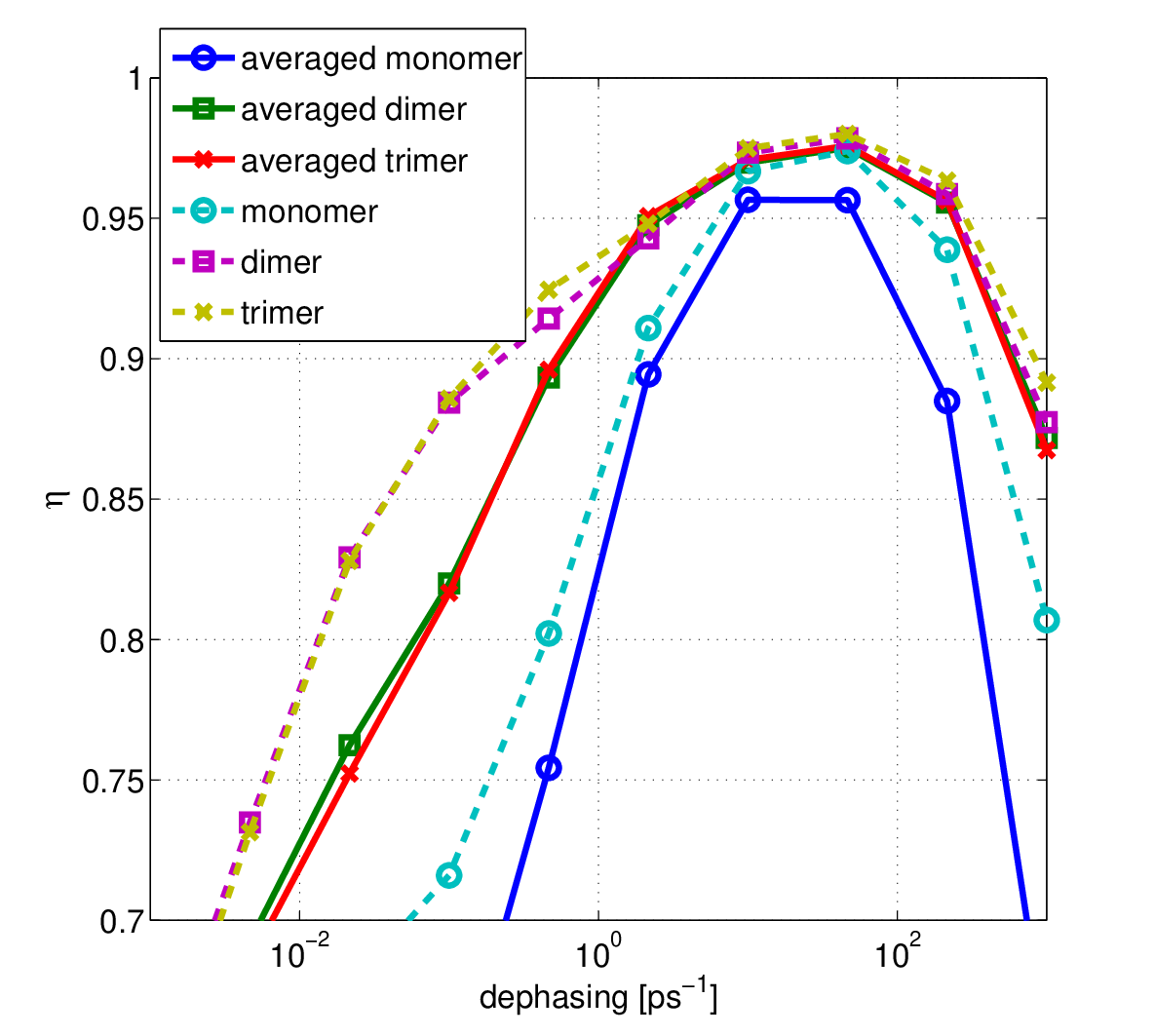}
\caption{Efficiency $\eta$ for different topologies (monomer, dimer, trimer)
in the presence of static disorder, varying the dephasing rate $\gamma_\phi$. Solid
lines represent the averaged value, while dashed lines represent the result
for the non-disordered Hamiltonian that we use in this work.}
\label{Fig.:disorder}
\end{figure}

 The results we obtain in this paper are robust with respect to
the introduction of static disorder.
In Fig.\ref{Fig.:disorder} we add the effects of static disorder (on site-energy values) and
compare the efficiency for different numbers of monomers.
The strength of disorder is taken to be that reported in \cite{disorder}.
The qualitative behavior of the efficiency with disorder is the
same as that without.  The monomer is more affected by disorder
than the dimer and trimer, indicating greater robustness for the
more complicated geometries. This is consistent with the fact
that fluctuations due to disorder are stronger for systems of smaller
size.

\begin{figure}[h!]
\fbox{\includegraphics[width=0.45\textwidth,clip]{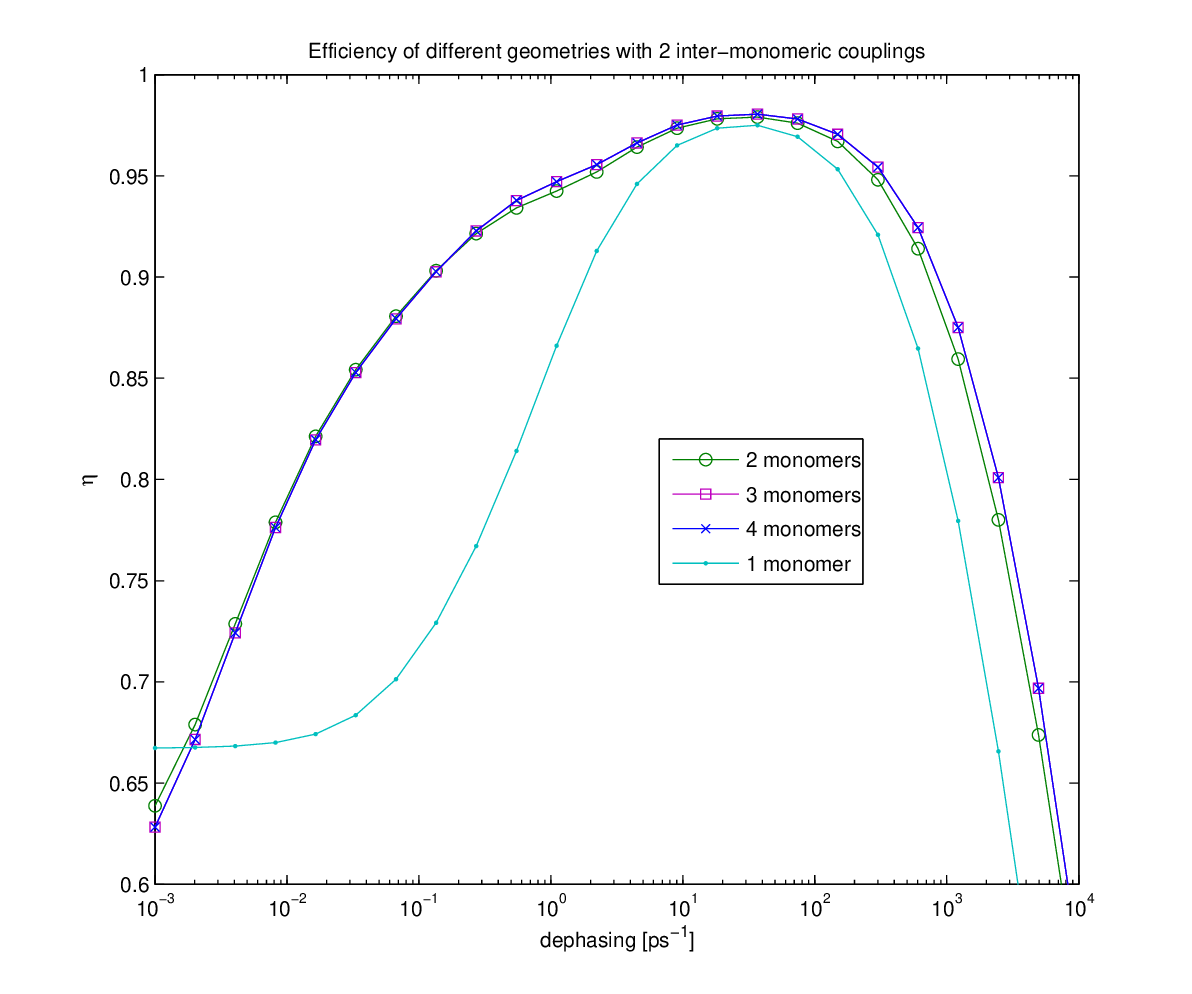}
\includegraphics[width=0.45\textwidth,clip]{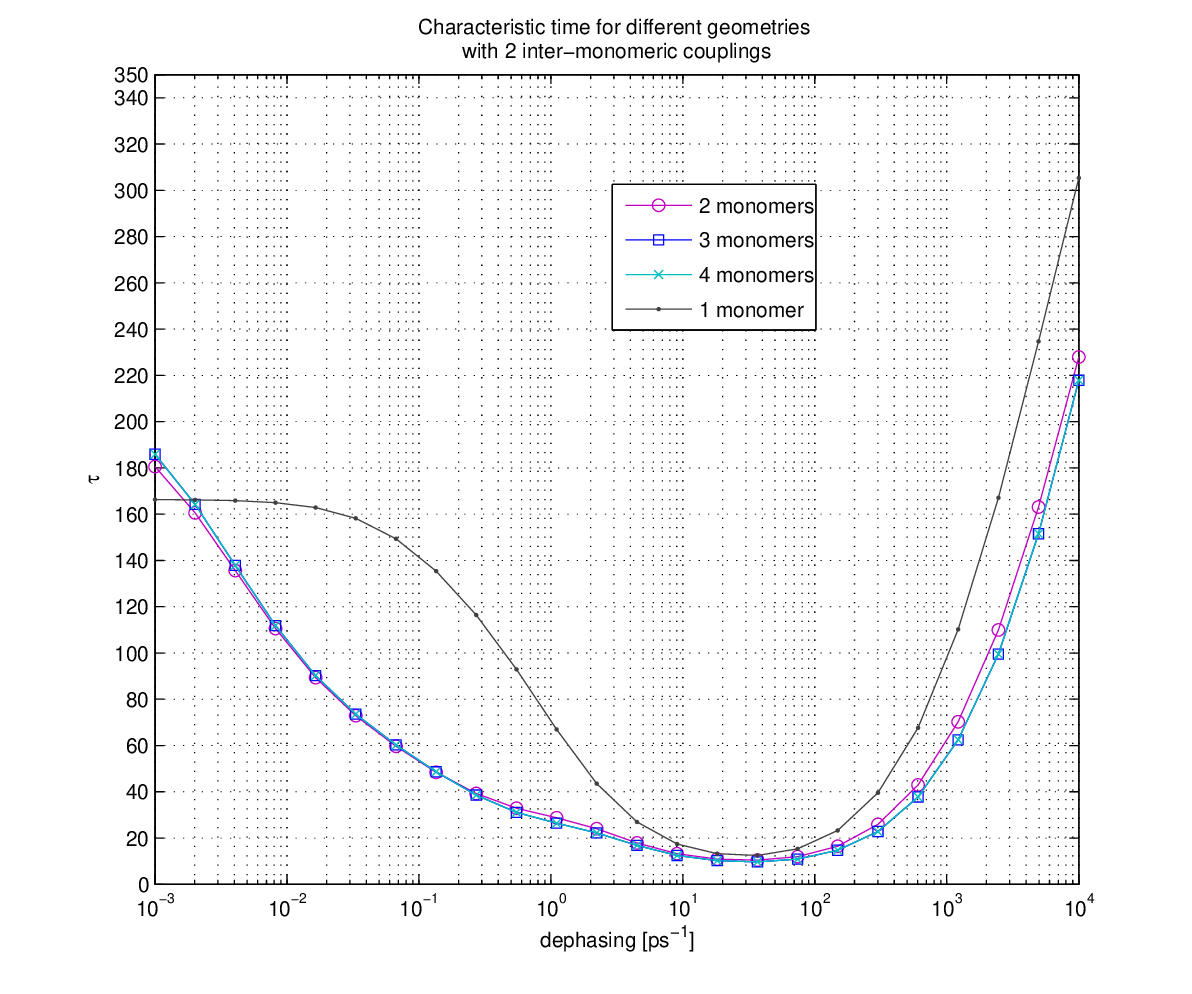}}
\caption{Efficiency $\eta$ (left) and average transfer time $\tau$ ($ps$) (right)for the different topologies (monomer, dimer, trimer and complex with 4 monomers)
with  {\bf two} intra-monomeric coupling for various values of the dephasing rate $\gamma_\phi$. Initial
state: highest energy eigenstate of the given structure.}
 \label{Fig.: donorEtaTauC189}
\end{figure}

Differences between the various geometries can be observed when the inter-monomeric
coupling is supposed to be present between the b601 Chl in one monomer and both b608 and b609
in the neighbouring monomer, see fig. \ref{Fig.: donorEtaTauC189}.
 The efficiency of the clusters of monomers benefits from this
kind of coupling: $\eta$ is enhanced with respect the single monomer case
over a wide range of dephasing values.
Moreover, while the dimer is slightly less efficient than
the trimer and the tetramer structures that have a higher number of traps,
our simulations suggest that there is no advantage in adding more than three
subunits: the trimer behaves
just as well as the tetramer.
This results could be an indication for a functional selection of the
trimeric configuration with respect to the other ones: the trimer could be the result of
an optimization with respect to the ``cost" of the structure.

\begin{figure}[h!]
\fbox{\includegraphics[width=0.32\textwidth,clip]{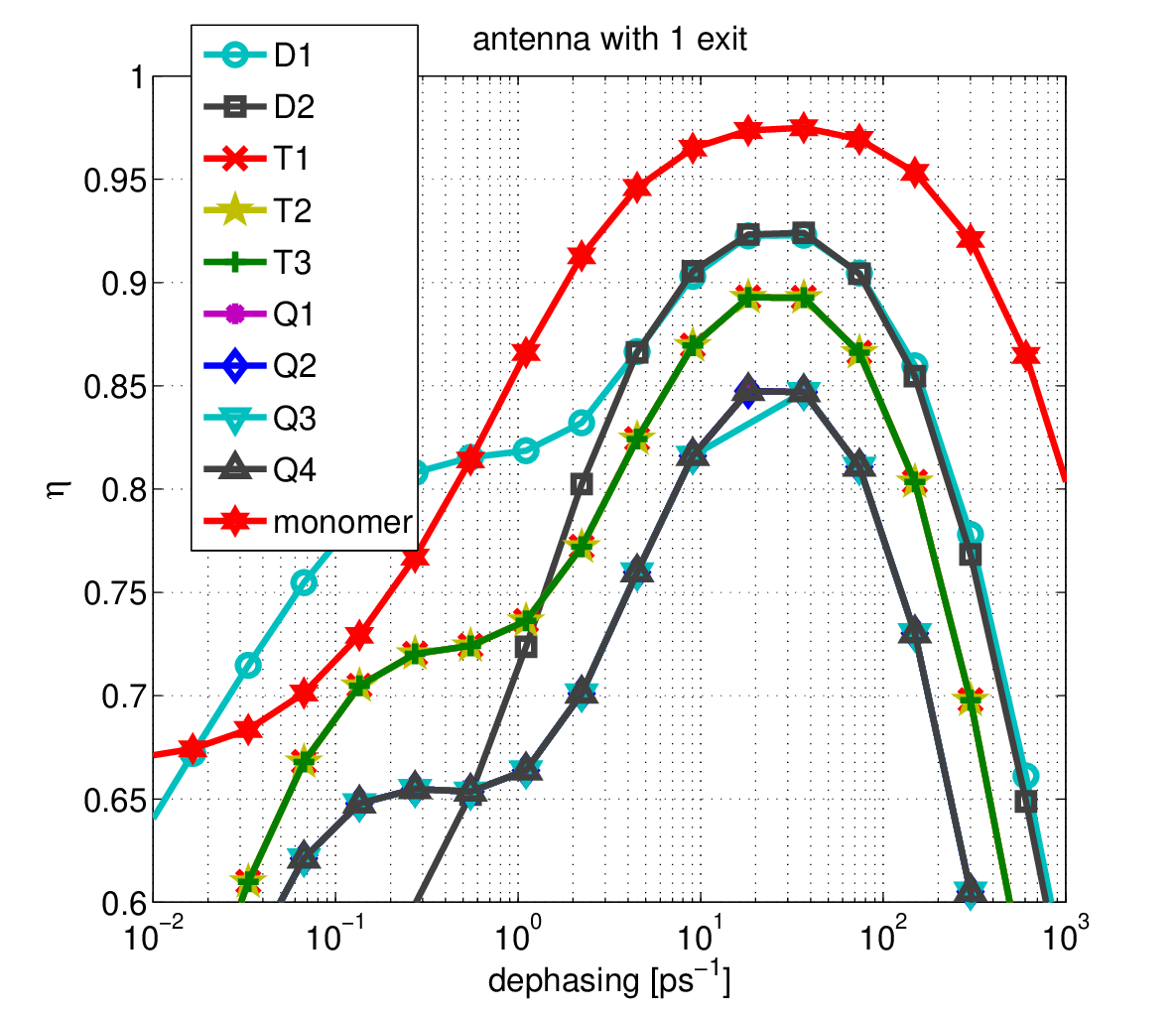}
\includegraphics[width=0.32\textwidth,clip]{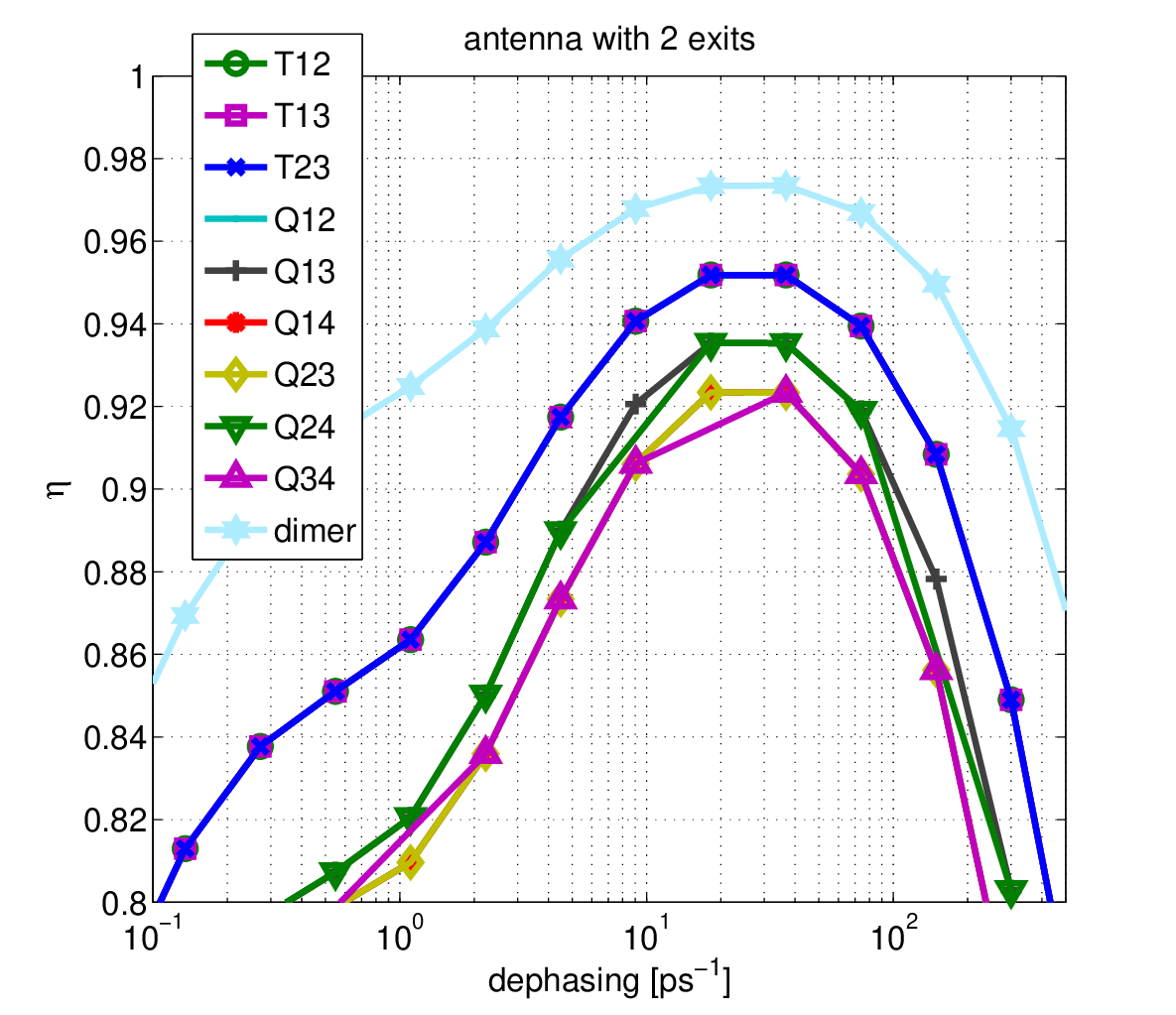}
\includegraphics[width=0.32\textwidth,clip]{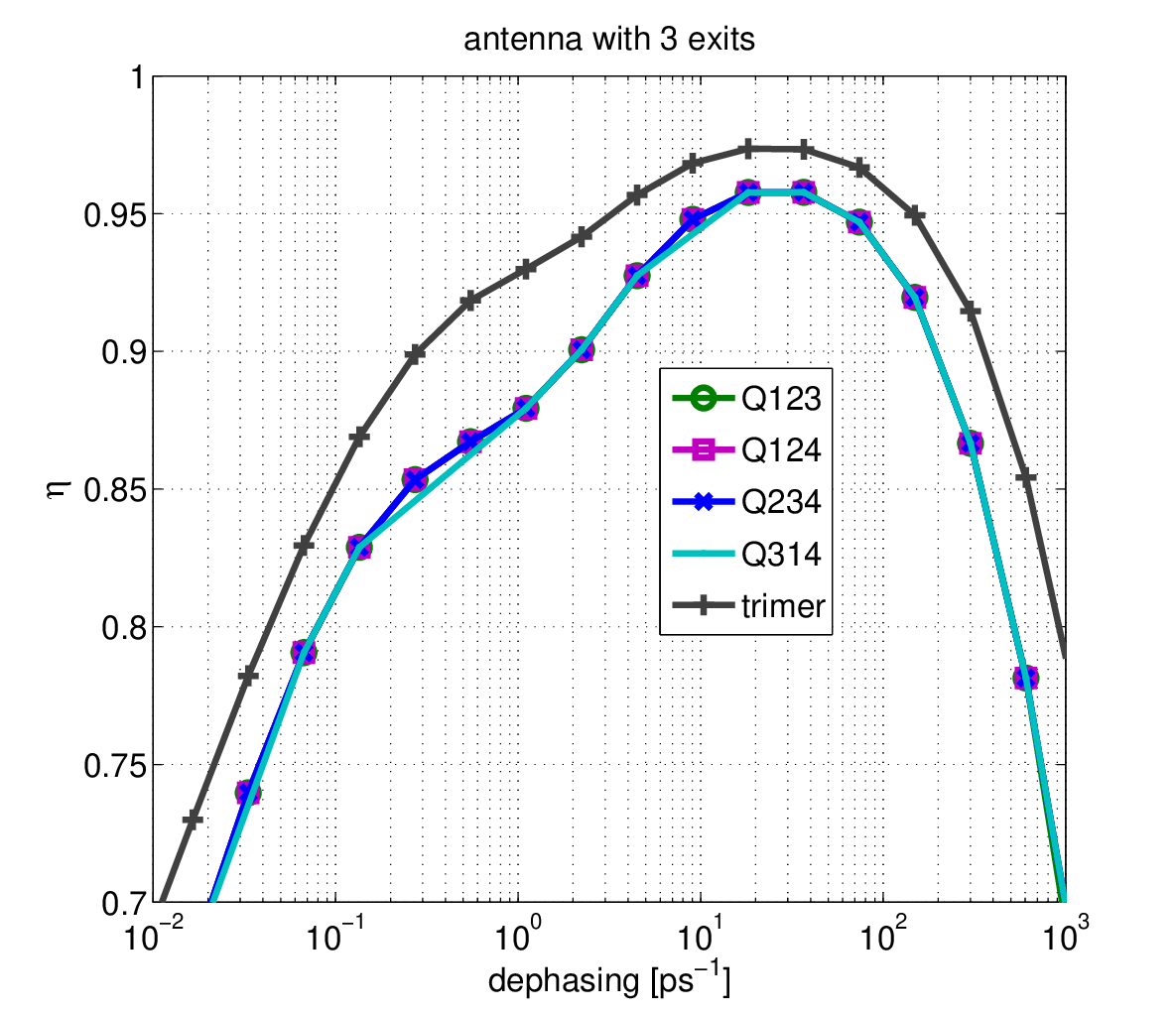}}
\caption{Efficiency of the transport for dimer, trimer and tetramer clusters, with one inter-monomeric
coupling, acting as ANTENNAS In the legend Q23 means that we consider a tetrameric complex where the
exciton is captured in the first monomer, and can exit only from the II and III monomers. Analoguosly
for the trimer (T) and the dimer (D).}
\label{fig:W3vs4Eta}
\end{figure}

Interesting differences between the behaviour of the
different structures appear when one considers a
variable number of active sinks. In
Fig. \ref{fig:W3vs4Eta} we compare the
transport in monomeric and dimeric, trimeric and tetrameric antennas  with only one inter-monomeric coupling,
and with different numbers of sinks attached to the available monomeric subunits.
The initial state in these simulations is always the highest excited
state of the structure.
 The main feature here is that the efficiency at physiological temperatures is
always greater for the structure with a number of sinks equal to the number of monomers.
For example, with a single sink attached, the monomer is more efficient than the dimer,
trimer and tetramer. This behaviour is reasonable since the exciton is initially delocalized over the
whole structure and in particular over those monomers that do not have any sink attached.
A qualitative explanation of the relative behavior
for the different configurations is based on the competition
between an enhancement of the efficiency due to a greater
number of sinks, and the depletion of efficiency due to
a greater number of chromophores, where the exciton can
delocalize and dissipate in the bath, causing
a slow down in the funneling process.
As already noticed, Fig. \ref{Fig.: donorEtaTauC19}, the trimer saturates the enhancement of
efficiency due to the number of exits: the tetrameric complex
with four sinks and the trimer complex with three sinks have the same
efficiency.

\subsection{The wire configuration}
We now analyze the multi-monomer structures when
they are used as ``wires".
As noted above, the LHCII complexes can function
as connecting structures between
different units of the PSII complex. In the following
we analyze this case by fixing as the initial state
of the dynamics the ground state $\ket{E_1}$ of one single
monomer in the cluster and by connecting a sink
{\em to each of the other monomers} in the complex.

\begin{figure}[h!]
\fbox{\includegraphics[width=0.45\textwidth,clip]{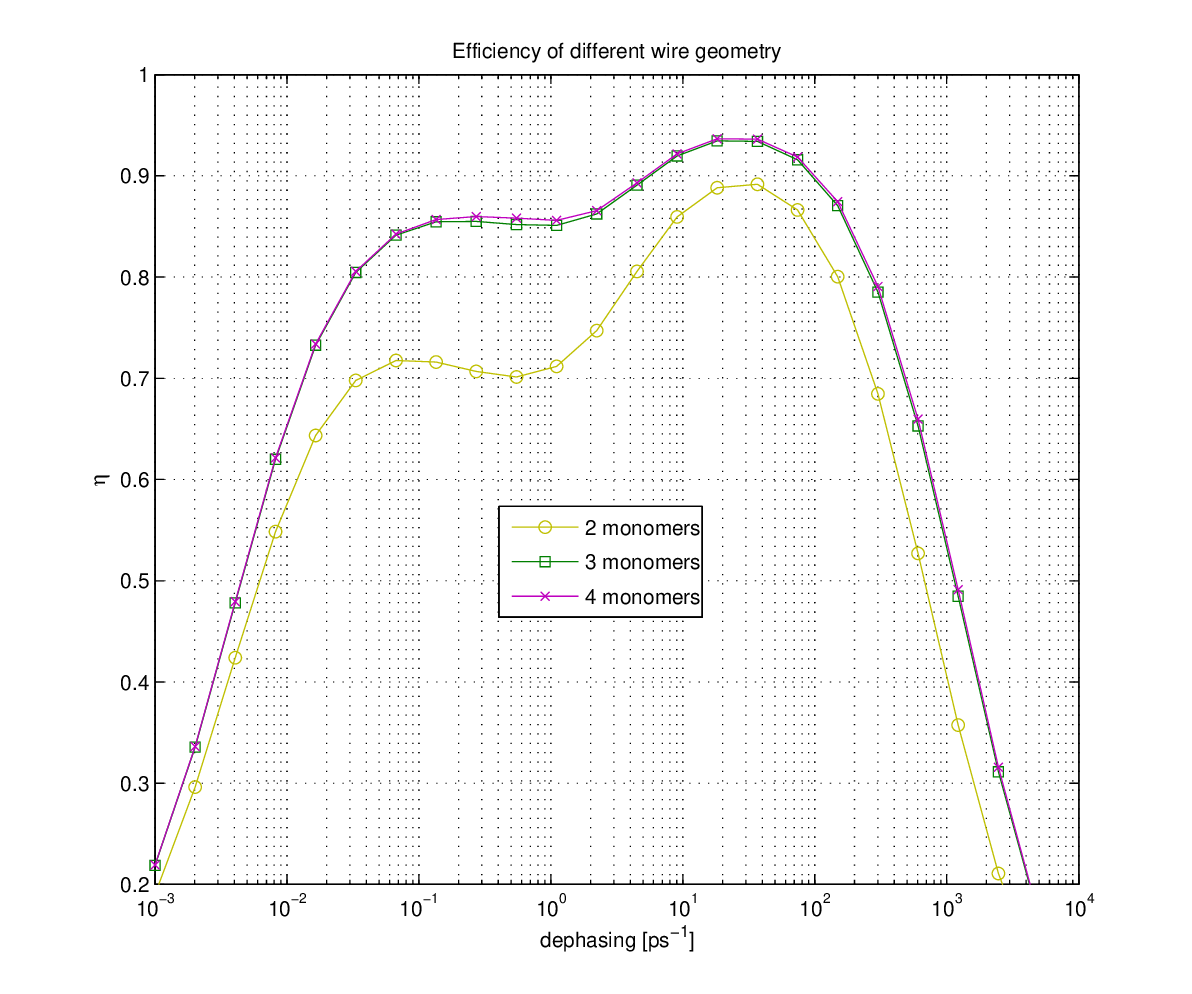}
\includegraphics[width=0.45\textwidth,clip]{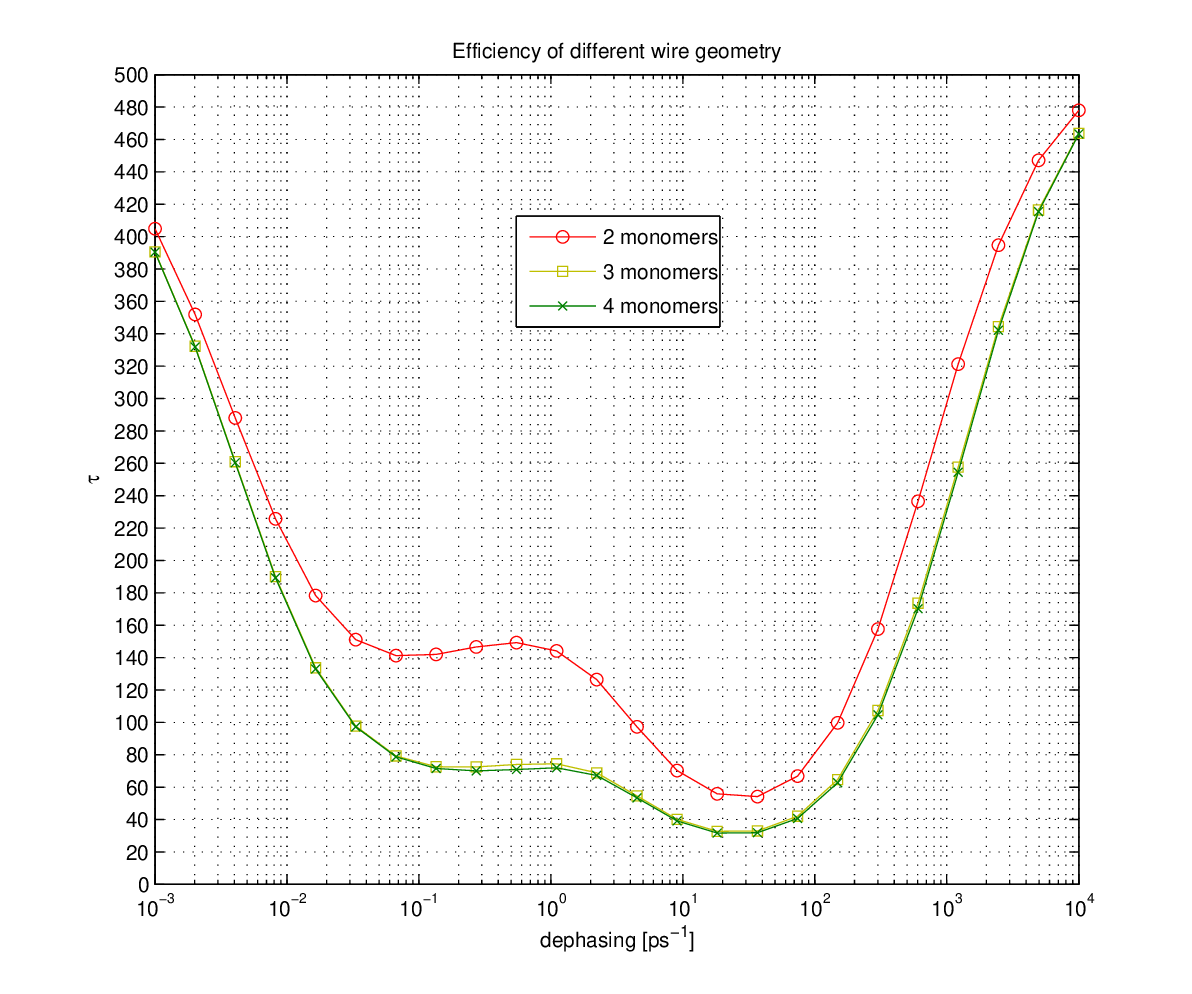}}
\caption{Efficiency $\eta$ (left) and average transfer time $\tau$ ($ps$) (right) for the different topologies (dimer, trimer and complex with 4 monomers)
with {\bf single} intra-monomeric coupling for various values of the dephasing rate $\gamma_\phi$. Initial
state $\ket{E_1}$, i.e. ground state of a single monomer, localized on the "acceptor" sites $a610,a611,a612$
.} \label{Fig.: wireEtaTau}
\end{figure}

%\begin{figure}
%\includegraphics[scale=1]{wire_3vs4_diffExit_eta.eps}
%\caption{Efficiency of the transport for dimer, trimer and tetramer clusters, with one inter-monomeric
%coupling, acting as wires. In the legend Q23 means that we consider a tetrameric complex where the
%exciton is captures in the first monomer, and can exit only from the II and III monomers. Analoguosly
%for the trimer (T) and the dimer (D).}
%\label{fig:W3vs4Eta}
%\end{figure}

%\begin{figure}
%\includegraphics[scale=1]{wire_3vs4_diffExit_tau.eps}
%\caption{Characteristic time of the transport for dimer, trimer and tetramer clusters, with one inter-
%monomeric coupling, acting as wires. In the legend Q23 means that we consider a tetrameric complex where
%the exciton is captures in the first monomer, and can exit only from the II and III monomers. Analoguosly
%for the trimer (T) and the dimer (D).}
%\label{fig:W3vs4Tau}
%\end{figure}

Fig. \ref{Fig.: wireEtaTau} shows the efficiency
and the characteristic time of the dimer, the trimer and of a cluster of
four monomers coupled together with only one inter-monomer connection.
The simulation suggests again a special role played by the trimer.
It is significantly more efficient than the dimer, but almost indistinguishable
from the cluster of four monomers.  
In the wire configuration, the exciton passing
through the dimer has only one trapping site that it can
reach, while in the trimer, tetramer, and higher
order polymers, the exciton entering at the lowest
energy site has two trapping sites that it can reach --
those on the two monomers adjacent to the entrance site.
Consequently, the efficiency of trapping is higher
for the trimer and tetramer than for the dimer:
two traps are better than one.
Moreover, the dynamics for passing from the entrance
site to those two adjacent traps are identical for the
trimer, tetramer, and higher order polymers.
Consequently, the efficiency profiles for the trimer, tetramer,
and higher order polymers are identical.

\begin{figure}[h!]
\includegraphics[width=0.5\textwidth,clip]{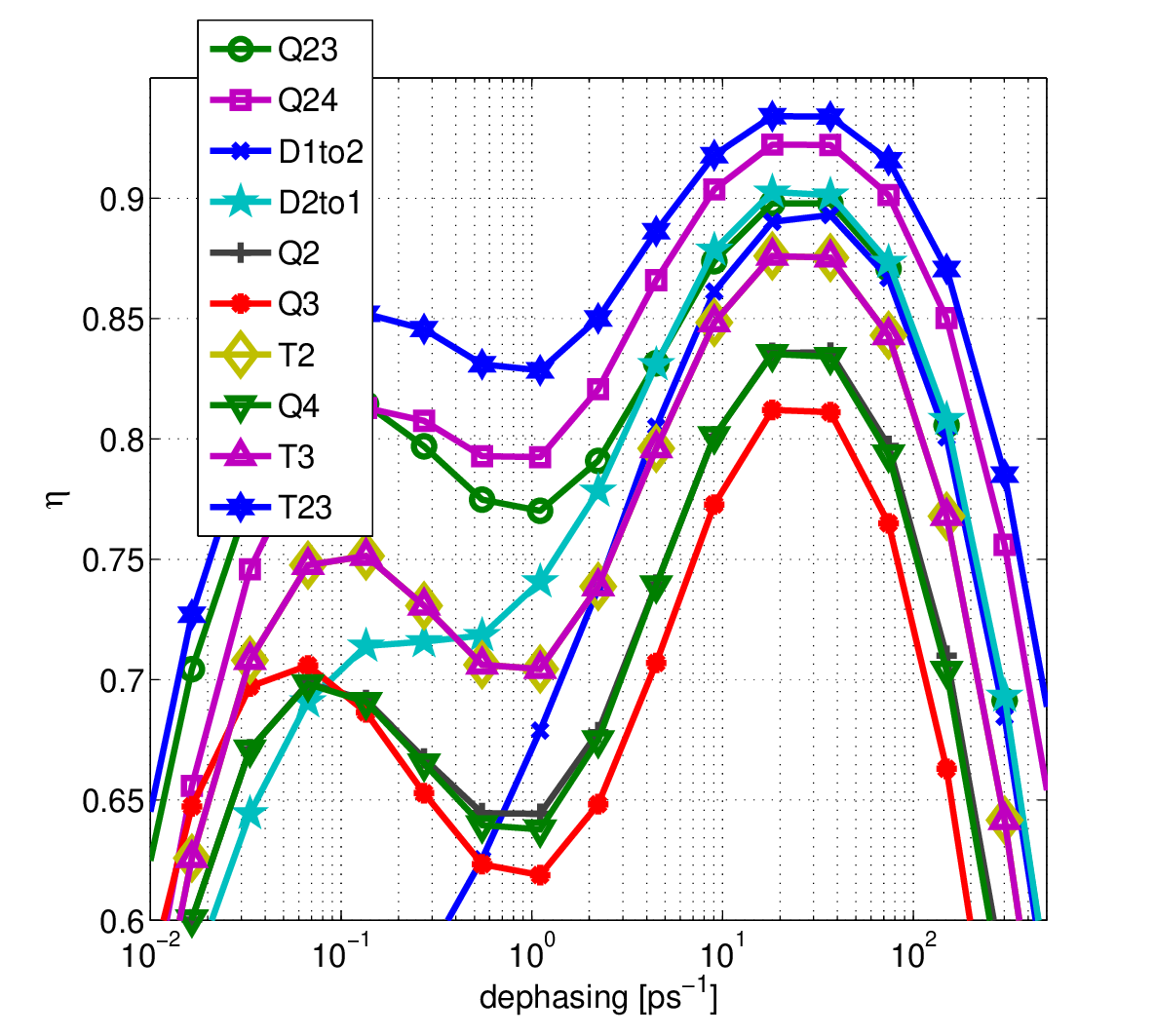}
\caption{Efficiency of the transport for dimer, trimer and tetramer clusters, with one inter-monomeric
coupling, acting as WIRES. In the legend Q23 means that we consider a tetrameric complex where the
exciton is captured in the first monomer, and can exit only from the II and III monomers. Analoguosly
for the trimer (T) and the dimer (D).}
\label{Fig.: wires_fullComp_eta}
\end{figure}

We finally describe how the number of attached sinks modifies the overall efficiency of the different quantum wires.
In Fig. \ref{Fig.: wires_fullComp_eta} the simulations refer to a situation where the initial
state of the structure is the ground-state $\ket{E_1}$ localized in the $a610,a611,a612$ Chls of a single monomer
of the structure.
 Here again we see a behaviour similar to the antenna confirguration case: at fixed number of sinks attached
the efficiency is higher for those structures with a smaller number of chromophores .
The trimer is again a limiting case, it is always more or as efficient as the tetramer, and, as already shown in fig. (\ref{Fig.: wireEtaTau}) the trimer with two sinks
has the same performance of the four-monomer structure with three sinks.
An analogue comparison holds true for  the dimeric vs trimeric cluster.
When only one sink is active, at the relevant values of dephasing --
i.e. where the efficiency is maximal -- the dimer can perform slightly better
than the trimer.
From our simulations it is also evident that, within our assumptions for the inter-monomer links, the
various structures show a directionality of transport.
For example the efficiency of the dimer changes significantly over a wide range of dephasing values
depending on the direction of the energy flow.

\section{Conclusion}

 We have analyzed excitonic transport in the primary component of the photosynthetic
apparatus in green plants: LHCII. By means of a simple decoherence model
(Haken-Strobl),  the analysis shows that an exciton initially
localized on a single chromophore moves through the LCHII photocomplex in
a two-step process which clearly signaled by a quantum information motivated measure of delocalization.
First, over the timescale of a few hundreds of picosecond,
the exciton spreads coherently to neighboring chromophores. This coherent spreading
exhibits rapid quantum oscillations and entanglement.
Second, as the environment decoheres the exciton's position,
the exciton diffuses semi-coherently throughout the complex over
a timescale of ten to twenty picoseconds.
Although the Haken-Strobl model does not include relaxation,
we expect this two-step, coherent--semicoherent model to hold
for more detailed models of the dynamics as well, for example,
in the full hierarchy approach of \cite{IshiFleming09}.

 Detailed comparison of different measures of correlations also allows us to
identify how the two main downhill relaxation pathways unveiled by the experiments
(stromal-stromal, lumenal-stromal) can be dynamically described in terms of correlations pathways:
the correlations among subsytems of pigments are mostly established on the stromal side
of the monomeric complex even when the excitation is initially localized on the lumenal side.
This behaviour has important consequences when the efficiency of the transport is considered:
the stromal pathway is in general more efficient than the lumenal one.

In general, the analysis shows that even in the absence of relaxation, pure dephasing
induces effective transport down the LHCII energy funnel.  The transport
is efficient and robust in the presence of static disorder,
and exhibits the characteristic signature of characteristic
environmentally assisted quantum transport (ENAQT) -- low
efficiency at low temperature due to transient localization,
followed by a robust maximum efficiency at physiological temperature,
with a falling off of efficiency at very high temperature.
 In the second part of the paper, we compared the efficiency of
transport through LHCII structures with different topologies (monomers, dimers, trimers, and tetramers)
and different configurations (antenna and wire).
We find that the efficiency of transport increases as the
number of subunits increases, saturating at the level of
the trimer. These results provide evidence for the functional selection of the
trimeric configuration.

\begin{acknowledgments}
PG would like to thank A. Ishizaki, L. Valkunas and T. Mancal for useful discussions and suggestions.\\
PZ acknowledges support  from NSF grants PHY-803304,  PHY-0969969 and DMR-0804914.
\end{acknowledgments}

\end{document}